\documentclass[a4paper,10pt,twoside]{cpc-hepnp}
\usepackage{CJK,upgreek,fancyhdr}
\usepackage{multicol}
\usepackage{graphicx}
\usepackage{booktabs}
\usepackage{amssymb,bm,mathrsfs,bbm,amscd}
\usepackage[tbtags]{amsmath}
\usepackage{lastpage}
\usepackage{color}
\usepackage{amsmath}
\usepackage{epstopdf}
\usepackage{amsfonts}
\usepackage{cancel}

\newcommand{\ppbar}{p \overline{p}}

\newcommand{\jpsi}{J/\psi}

\begin{document}
\begin{CJK*}{GB}{gbsn}

%\fancyhead[c]{\small Chinese Physics C~~~Vol. xx, No. x (201x) xxxxxx}
%\fancyfoot[C]{\small 010201-\thepage}
%
%\footnotetext[0]{Received 31 June 2015}

\title{Exotic hadron bound state production at hadron colliders
\thanks{Supported by Natural Science Foundation of Shandong Province (ZR2014AM016, ZR2016AM16) and National Natural Science Foundation of China (11275115, 11325525, 11635009)}}

\author{%
      Yi Jin (œðÒã)$^{1;1)}$\email{ss\_jiny@ujn.edu.cn}%
\quad Shi-Yuan Li (ÀîÊÀÔš)$^{2;2)}$\email{lishy@sdu.edu.cn}%
\quad Yan-Rui Liu (ÁõÑÔÈñ)$^{2}$
\\
\quad Lu Meng (ÃÏèŽ)$^{2}$
\quad Zong-Guo Si (ËŸ×Ú¹ú)$^{2,3}$
\quad Xiao-Feng Zhang (ÕÅÏþ·å)$^{2}$
}
\maketitle

\address{%
$^1$ School of Physics and Technology, University of Jinan, Jinan 250022, China\\
$^2$ School of Physics, Shandong University, Jinan 250100, China\\
$^3$ CCEPP, Institute of High Energy Physics, Beijing 100049, China\\
}

\begin{abstract}
	The non-relativistic wave function framework is applied to study  the production and decay of the exotic hadrons which can be effectively described as bound states of other hadrons.
	Employing the factorized formulation, we investigate the  production  of  exotic hadrons in the multiproduction processes  at high energy hadronic colliders with the help of event  generators.	This study provides crucial information for the measurements of the relevant exotic hadrons.
	% 	
%The abstract should summarize the context, content
%and conclusions of the paper in less than 200 words. It should
%not contain any references or displayed equations. Typeset the
%abstract in 8~pt Roman, with
%an indentation of 1.5 pica on the left and right margins.
\end{abstract}



\footnotetext[0]{\hspace*{-3mm}\raisebox{0.3ex}{$\scriptstyle\copyright$}2013
Chinese Physical Society and the Institute of High Energy Physics
of the Chinese Academy of Sciences and the Institute
of Modern Physics of the Chinese Academy of Sciences and IOP Publishing Ltd}%

\begin{multicols}{2}

\section{Introduction}\label{sec1}

 Recently, more and more new exotic hadron states,  {\it e.g.}, the  XYZ mesons, have been observed. They are assumed as multiquark states and/or as bound states of other ingredient hadrons in lots of theoretical investigations \cite{Brambilla:2010cs,Yuan:2014rta,Swanson:2015wgq,Chen:2016qju}. Such investigations can be done not only via their decay processes, where the branching ratios and distributions of the decay products can be studied, but also via their production processes (from the decay of the heavier particles or directly from the multiproduction processes). In both cases, the more complex the process is, the more information of inner structure can be drawn. In general, the production processes are more complex.

At the same time, the studies on the production processes also provide the information of the cross section, rapidity and transverse momentum distributions,  {\it etc.}, of the relevant particles on a specific collider, which can help the experimentalists to set the proper triggers and cutoffs for the measurements \cite{Jin:2014nva}. A good  example is  Large hadron Collider (LHC) where various detectors cover a large rapidity range. They can be used to study exotic hadron production in B-decays \cite{Kahana:2015tkb}, as well as direct production in the multiproduction process of high energy hadronic collisions and  nuclear collisions. So the distributions mentioned above are crucial for the studies on exotic hadrons with a specific detector at LHC.

Furthermore, the direct production of the exotic hadrons in the multiproduction process of high energy scattering can set a crucial point for the understanding of the hadronization mechanism. Since the exotic hadrons always refer to the states with more than three constituent quarks (here we do not discuss hybrids or glueballs), one of the feasible ways for understanding their production mechanism is to employ the combination model \cite{Jin:2015mla} to combine the necessary constituent quarks into the relevant hadron. However, in any hadronization process, as pointed in Refs. \cite{Han:2009jw,Jin:2013bra}, the produced color-singlet (anti)quark system eventually transits to various hadron states (the mesons, baryons and beyond) with the total probability exactly 1:
\begin{equation}\label{ueq1}
	\sum_h |U_{hq}|^2 = \sum_h |\langle h|U|q\rangle|^2 =\langle q|U{^+}U|q\rangle=1.
\end{equation}
Here we introduce the unitary time-evolution operator $U$ to describe the hadronization process. For the quark state $|q\rangle$ and the corresponding hadron state $|h\rangle$, the matrix element $U_{hq}=\langle h|U|q\rangle$ describes the transition amplitude. $U_{hq}$ is determined by (low energy) Chromodynamics (QCD) but beyond the present approach of calculation. This leaves the space for various hadronization models to mimic the transition process.
The unitary operator $U$ reflects the fact that there are no free quarks in the final states of any high energy process,  {\it e.g.}, the so-called quark confinement. The introduction of multiquark states sets a challenge for the hadronization models dealing with the transition from color-singlet (anti)quark system to the hadron system.

As a matter of fact from  experiments, the production of general mesons and baryons is dominant, {\it i.e.},
\begin{equation}
	\label{ueqless1} \sum_{h=B,\bar B,M} |\langle h|U|q\rangle|^2 \sim 1-\varepsilon,
\end{equation}
here $B$, $\bar B$ and $M$ denote baryon, antibaryon and meson, respectively. If the exotic hadrons are produced,  $\varepsilon$ could be a small but non-vanishing value. Since the production rate is proportional to the quark density to the power of constituent quark number in the hadron, in the cases of large number of quarks produced such as those in high energy nuclear collisions, the more constituent quarks a hadron contains, the larger production rate one gets. So to regain the unitarity, one needs the special `combination function' which reflects the confinement and may be related with the whole system rather than the several quarks to be combined into a specific hadron \cite{Han:2009jw,Jin:2013bra}. Since the present knowledge is not enough to judge how many kinds of multiquark states there are and how they `share' the total probability of $\varepsilon$, one can not predict the production rate of a specific multiquark state. What we can suspect, though, seems that if there are a lot of kinds of multiquark hadrons, each only shares a small part of the small $\varepsilon$. So the production rate of each is almost vanishing.

However, if one of the exotic states is the bound state of other hadrons,  {\it i.e.}, its production can be taken as from the combination of mesons and/or (anti)baryons, there is generally no straightforward unitarity constraint as above on its production rate. Unlike quarks, hadrons are not confined. They can be either free, or bound with other hadron(s), even lepton(s) ({\it e.g.}, hydrogen atom). As a matter of fact, in the cases that the number of produced hadrons are large, such as in high energy ion collisions, the familiar hadron bound states, such as deuterons, $\alpha$ particles,  {\it etc.}, have been observed.
So it is natural to investigate the productions of the `hadronic molecule' relevant to exotic states. The measurements and explanations of large production rate of X(3872)
\cite{Aaltonen:2009vj, Abulencia:2006ma, Kerzel:2006ks, Abazov:2004kp, Aaij:2011sn, Chatrchyan:2013cld, Kordas:2016ebh, Bignamini:2009sk, Artoisenet:2009wk, Bignamini:2009fn, Artoisenet:2010uu, Guo:2013ufa, Guerrieri:2014gfa}
and X(5568) \cite{D0:2016mwd, Jin:2016cpv, Aaij:2016iev, CMS:2016fvl} are also good examples.
Their large production rates lead to insights on the investigation of their structure and production mechanism, especially the colour and spin structures \cite{Artoisenet:2009wk,lj}.

For those XYZ states possibly considered as hadronic molecules, we can describe them with the framework of various Non-Relativistic (NR) effective theories, especially the NR wave function method, and concentrate on their inclusive production. 
The ingredient hadrons in the hadronic molecule are loosely bound, hence the relative momentum between them is fairly small with respective to the hadron mass (almost of the order of charm quark), so the hadronic molecule is in principle a NR system.

Besides the application of the investigation on positronium, the NR wave function is also used for the heavy quarkonium production and decay, generally referred to as the `color-singlet model'.

The Non-Relativistic Quantum Chromodynamics (NRQCD) implies that the quark pair in color octet could also transfer into a color-singlet hadron, which is referred to as the color-octet model and used to explain the production rate and transverse momentum of prompt quarkonium in hadronic collisions.  But for the case of hadron as basic degree of freedom, there is no problem of color confinement because every object is color-singlet. Hence the relevant complexity \cite{Li:1999ar,Han:2006vi} is eliminated. If the system of bound hadrons is properly modeled and the NR wave function is obtained, the NR wave function framework can be used for various decay and production processes of hadronic molecules.

This method has been applied to the near threshold enhancements of mass spectra in the  $J/\Psi \to \gamma \ppbar$ and $J/\Psi \to p \bar \Lambda K^-$ channels of the production of bound states X($\ppbar$) and X($ p\bar \Lambda$) \cite{noteictp}. In that note, the decay to the corresponding ingredient hadrons is described by an effective Lagrangian. However, for the direct production of the mesons and baryons at high energy hadronic collisions, it is impossible to construct the effective Lagrangian. In Ref. \cite{noteictp}, we suggested to employ the general event generator to extract the cross section of the ingredient hadrons.
%It is the task of this paper to investigate the details.
In fact, this is one of the advantages of the NR wave function framework. One can expand the amplitude with respective to the relative momentum between the ingredients because it is relatively small. Thus the cross section is factorized. Only the cross section (rather than the  amplitude) of the ingredient hadron production is needed, which can be fixed by fitting the correlation data  of the relevant hadrons.

In the next section, we will review and list the  formulations for the calculation of hadronic molecule production rate in the NR wave function framework, taking $pp$ scattering at LHC, say $pp \to A+B + X \to H(A,B) +X$, as an example. Based on  these formulations,  only the NR wave function  (and/or its derivatives) at the origin 
and correspondingly, only the square of the absolute value of the production amplitude  (and/or its derivatives) of ingredient hadron A and B, are  relevant. This means that  the details of  the structure of the bound state and the production of the ingredient hadron in all the phase space, are not fully used.
So this framework provide the benchmark of the information  least sensitive to the models and details of the strong dynamics in the momentum scale of several MeV to several ten MeV.  At the same time, if the universality of the wave function is confirmed, the factorization is well established.
In Section 3, the calculation of the cross section of the ingredient hadron pair of $A$ and $B$ will be investigated, extrapolating to the vanishing relative momentum. We make a comparison with present approach in literatures \cite{Bignamini:2009sk, Artoisenet:2009wk, Bignamini:2009fn, Artoisenet:2010uu, Guo:2013ufa, Guerrieri:2014gfa}.
Then the rapidity and transverse momentum distributions of a series of bound states like $K\bar{K}$, $D^*N$, $D\bar{D}^*$ (X(3872)), $\Lambda_c\bar{\Lambda}_c$, $\Sigma_c\bar{D}^*$,
%$B^0_s \pi^{\pm}$ (X(5568)) and $D^{\pm}_s \pi^{\pm}$
X(5568) and $X_c$ \cite{D0:2016mwd, Jin:2016cpv}, {\it etc.}, are taken as illustrations in Section 4, besides further discussions are given.

\section{Non-Relativistic wave function formulation for $pp \to A+B + X \to H(A,B) +X$}\label{sec2}

\begin{figure}
	\centering
	\vspace*{-1.5cm}
	\hspace*{-1.5cm}\includegraphics[scale=1.0]{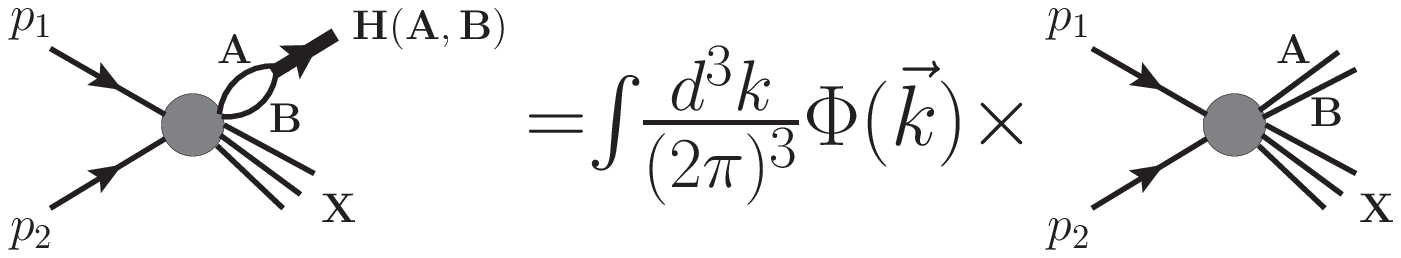}
	\vspace*{-4.0cm}
	\caption{The process $p(p_1)p(p_2) \to A(p_A)+B(p_B) + X \to H(A,B)(P_H) +X$.
		\label{feygl}}
\end{figure}

For the case that a bound state $H$  is well described  by the two ingredients $A$ and $B$, the only difference between the amplitude of the bound state and that of the free particles lies in that the wave function rather than the plane-waves of free particles is adopted to describe the bound state. The process $pp \to A+B + X \to H(A,B) +X$ is illustrated in Fig.~\ref{feygl}, and the corresponding invariant amplitude is:
\begin{equation}
	\label{startpoint}
	\begin{array}{l l l}
		A_{inv} & = & \langle H(A,B), X|\hat{T}|p p \rangle  \\
		& = & \frac{1}{\sqrt{\frac{{{m}_{A}}{{m}_{B}}}{{{m}_{A}}+{{m}_{B}}}}} \int \frac{d^3 k}{(2\pi)^3} \Phi(\vec k)
		\langle A|\langle B|\langle X|\hat{T}|pp\rangle  \\
		& = & \frac{1}{\sqrt{\frac{{{m}_{A}}{{m}_{B}}}{{{m}_{A}}+{{m}_{B}}}}} \int \frac{d^3 k}{(2\pi)^3} \Phi(\vec k)
		{\bold M}(\vec k).
		%\end{eqnarray}
	\end{array}
\end{equation}
This formulation is valid in the rest frame of $H(A,B)$, the bound state of the ingredient hadrons $A$ and $B$. In the above equation, $\vec k$ is the relative 3-momentum between $A$ and $B$  in the  rest frame of $H(A,B)$. ${\bold M}(\vec k)$ is the invariant amplitude  for the free (unbound) $A$ and $B$ production. The factor $1/ \sqrt{\frac{{{m}_{A}}{{m}_{B}}}{{{m}_{A}}+{{m}_{B}}}}$ comes from the normalization of the bound state to be $ 2 E_H V$ just as a single particle.  
%The normalization for the phase space wave function $\Phi(\vec k)$ is the same as the textbook of O. Nachtmann \cite{Nachtmann:1990ta}. 
 The normalization for the phase space wave function $\Phi(\vec k)$ is \cite{Nachtmann:1990ta} 
\begin{equation}
     \int \frac{d^3 k}{(2\pi)^3} |\Phi(\vec k)|^2 =1.
\end{equation} 
From the above equation, it is obvious that if we know the analytical form of the wave function as well as the free particle invariant amplitude, we can calculate the amplitude of the bound state simply by integrating the relative momentum $\vec k$.
In practice, certain decomposition and simplification  will be taken for concrete $^{2S+1}L_J$ state, as examples in the following. The resulting formulations are covariant.

\end{multicols}

%\ruleup
%\ruledown

\begin{multicols}{2}

The free particle invariant amplitude $\bold M (\vec{k}) $, as is shown by the part after the product sign in Fig.~\ref{feygl}, can be obtained from the corresponding Feynman diagrams with the Feynman rules from `standard' model or other effective theories (see the following discussions). The general form of this invariant amplitude is $B_j(p_B) \hat{O}_{ji}(P_H,\vec{k})  A_i(p_A)$.  In the following we take a spin-1/2 fermion-antifermion pair as an example to illustrate some details. In this case it is a Lorentz scalar $\bar u(p_B) \hat{O}(P_H,\vec{k})  v(p_A)$. Here $\hat{O}(P_H,\vec{k})$ represents a collective product of $\gamma-$matrices, spinors, {\it etc.}, {\it i.e.}, the internal lines, vertices, and other external lines except those of the $A$ and $B$. Hence in the rest frame of $H(A,B)$, the invariant amplitude of the bound state production is:

\begin{eqnarray}
	\label{basic}
	&& A(P_H, J J_z)=\frac{2}{\sqrt{M}}\sum_{S_z,L_z}\int \frac{d^3 k}{(2 \pi)^3} \Phi_{L,L_z}(\vec{k}) \nonumber \\
	&& \cdot \langle L,L_z;S, S_z|J, J_z\rangle \times \sum_{s,\bar{s}}\langle \frac12,s;\frac12,\bar s|S,S_z\rangle  \\
	&& \cdot \bar u(p_B, \bar s) \hat{O}(P_H,\vec{k}) v(p_A,s ), \nonumber
\end{eqnarray}
where $M$ is the mass of the bound state. The above equation has two C-G coefficients (spin 1/2~-~1/2 and spin-orbital angular momentum) for the specified angular momentum of $H$. The decomposition of  the two spin-1/2 addition is:

\begin{eqnarray}
	\label{spindec}
	&& \sum_{s, \bar s} \bar u(p_B, \bar s) \hat{O}(P_H,\vec{k})  v(p_A,s)  \langle \frac12,s;\frac12,\bar s|S,S_z\rangle \nonumber \\
	&& =\sum_{S S_z} \frac{1}{\sqrt{E_{A}+M/2}} \frac{1}{\sqrt{E_{B}+M/2}} \times  \\
	&& Tr\big {[}(M+\cancel{P}_H) \hat{O} \frac{1+\gamma_0}{4\sqrt{2}} \Pi_{S S_z}  \big {]}, \nonumber
\end{eqnarray}
where
\begin{equation}
	\Pi_{00}=-\gamma_5, ~\Pi_{1,S_z}=- \cancel{\epsilon}(S_z),
\end{equation}
with the polarization vector
\begin{equation}
	\epsilon^\mu (0)=(0,0,0,1), ~\epsilon^\mu (\pm 1)=(0,\mp 1,- i,0)/\sqrt{2}
\end{equation}
in the rest frame.
%with the value of
Since $\vec{k}$ is small in the rest frame, the above formulations can be expanded around $\vec{k}=0$ up to terms linear in $\vec{k}$ for the cases of S-wave and P-wave ($L=0, 1$) as:
\begin{equation}
	\label{expand}
	Tr \big {[}(M \hat{O}_0 + \cancel{k} \hat{O}_0 + M k_\mu \hat{O}^\mu) \frac{1+\gamma_0}{4 \sqrt{2}}\Pi_{SS_z} \big{]},
\end{equation}
where $\hat{O}_0=\hat{O}(\vec k=0)$ and
$\hat{O}^{\mu}=\frac{\partial}{\partial k_{\mu}}\hat{O}(\vec k)|_{\vec k=0}$.

For the S-wave case, $\langle L,L_z;S, S_z|J, J_z\rangle=\delta_{J,S}\delta_{J_z,S_z}$ in Eq. (\ref{basic}).
The wave function in Eq. (\ref{startpoint})
is symmetric and only the first part in Eq. (\ref{expand}) has contribution, hence

\begin{eqnarray}
	\label{swave}
	&& A^{L=0}_{J=S}(P_H)=\frac{2}{\sqrt{M}} \int \frac{d^3 k}{(2 \pi)^3} \Phi^{L=0}(\vec k) Tr \big {(}M \hat{O}_0 \frac{1+\gamma_0}{4 \sqrt{2}}\Pi_{S S_z} \big {)} \nonumber  \\
	&& =\frac{1}{2\sqrt{2}} \frac{1}{\sqrt{M}} \Psi^{L=0}(0) Tr \big {[} \hat{O}_0 (M+\cancel{P}_H)\Pi_{S S_z} \big {]},
\end{eqnarray}
where the coordinate space Schr\"{o}dinger S-wave function  $\Psi^{L=0}(0)=\frac{1}{\sqrt{4 \pi}} {\cal R}^{L=0}(0)$. The radial wave function $\cal R$ can be obtained when the interaction potential between the ingredients is determined.
For $L=1$, the wave function is anti-symmetric, the first part in Eq. (\ref{expand}) does not contribute, but the second and third term give, so the amplitude for the $^{2S+1}L_J=^1P_1$ bound state is:
\begin{equation}
	% A_{^1P_1}=\sqrt{\frac{1}{2}} \sqrt{\frac{2}{M}} \sqrt{\frac{3}{4 \pi}} {\cal R}'(0)
	A_{^1P_1}= \frac{1} {\sqrt{M}} \sqrt{\frac{3}{8 \pi}} {\cal R}'(0)
	Tr \Big {(} \cancel{\epsilon} \hat{O}_0 \frac{\cancel{P}_H}{M}+\epsilon^\mu \hat{O}_\mu \frac{M+\cancel{P}_H}{2} \Big {)}. %\gamma_5.
\end{equation}
The above formulations can all be found from the literature, {\it e.g.}, \cite{Kuhn:1979bb}, and are widely applied, {\it e.g.}, in quarkonium production.

As demonstrated above, by the recognition  that the relative momentum between the ingredients of the NR bound system is small, the expansion of the amplitude leads to the simplification of factorized formulation at cross section level.
We need
\begin{itemize}
	\item the production cross section rather than the amplitude of free particle with vanishing relative momentum,
	and the ingredient particles on shell, projected onto the special quantum number of the definite bound state (mainly the isospin and/or angular momentum state).
	\item Schr\"{o}dinger wave function to describe the bound state, which can be obtained from the potential  models of the relevant bound system.
\end{itemize}

The way to get the distribution of free  pair production sometimes can  rely on the effective theories based on  QCD  concept, to calculate the amplitude then the cross section.
For example, in \cite{noteictp}, the processes of $\jpsi$ decay into the baryon, anti-baryon and one or more pseudo-scalar particles, {\it e.g.}, kaons or pions,  with the pair of fermions forming  a bound state is studied. The relevant decay process $J/\Psi \to p \bar \Lambda K^-$ is described by a simple effective interaction Lagrangian.
We calculated  the partial width  of the process $J/\Psi \to p K^- \bar \Lambda $ to determine the coupling constant $G$. However, for very complex processes such as the production of hadrons in high energy  $pp$($\bar p$) collision, the effective Lagrangian can not be constructed. In this case, to employ the event generators to give the distribution is the only practical way. We will show that the generator can also be applicable to the P-wave case.

\section{Studying free hadron pair with event generators}\label{sec3}

The free pair cross section $p(p_1)p(p_2) \to A(p_A)+B(p_B) + X $ can be expressed as:
\begin{eqnarray}
	\label{int2}
	&& \frac{1}{N} \frac{d N}{d^3 P_H d^3 q} \propto \frac{1}{F} \sum_{j \neq A, B} \hspace{-0.67 cm} \int \prod \frac{d^3 p_j}{(2 \pi)^3 2 E_j} \nonumber \\
	&& \times \overline{ |\hat{O} |^2} (p_j, P_H=p_A+p_B, q=p_A-p_B)  \\
	&& \times (2 \pi)^4 \delta^{(4)}(P_{intial}-\sum_{j \neq A, B} p_j-p_A-p_B). \nonumber
\end{eqnarray}
Here the average is on various spin states, and the proper initial flux factor $1/F$ and phase space integral are needed. $\hat{O}$ is the amplitude of production of two free ingredient particles (with vanishing relative momentum and proper angular momentum state). It is not possible to be calculated directly with some effective quantum field theory/model when the initial state is (anti) protons and $A$ and $B$ are hadrons or clusters \cite{Jin:2016cpv}. However, it can be obtained with an event generator such as PYTHIA \cite{Pythia} or equivalently Shandong Quark Combination Model \cite{Jin:2010wg}, etc. for the case that $A$ and $B$ are both on shell.
% It is the advantage that in the above formulation only the on shell case is considered.
It is the advantage that in the above framework we employ, only the on shell case is considered, so that the numerical calculation with event generator is plausible.
The quantity of Eq. (\ref{int2}) describes the two hadrons/clusters ($A$ and $B$) correlation in the phase space. For the hadron case, by proper integral on components of $P_H$ and/or $q$, the resulting correlations can be directly compared with
data and serve for tuning the parameters.

Since the special physical picture of the non-relativistic framework, it is only valid in the rest frame of the two ingredient particles. One can define the following covariant space-like relative  momentum $\hat{q}$ as
\begin{equation}
	\hat{q}=(p_A-p_B) - \frac{(p_A-p_B) \cdot (p_A+p_B)}{(p_A+p_B)^2}(p_A+p_B).
\end{equation}
It is clear that in the rest frame of $A$ and $B$ ($H(A,B)$) where $\vec{p}_A+\vec{p}_B=0$, $\hat{q}=(0,\vec{k})$ and the $k=\sqrt{-\hat{q}^2}$ is exactly the absolute value of the 3-relative momentum $|\vec{p}_A-\vec{p}_B|$.

Employing the event generator, one gets
\begin{equation} \label{ext}
	\frac{1}{N} \frac{d N}{d^3 P_H d^3 \hat{q}}, \forall \hat{q},
\end{equation}
then extrapolates to the special case $k=0$. Numerically, one can take an average around $k=0$ for the above quantity.
Then we get, up to the kinematic factors as for the covariant form,
\begin{eqnarray}
	\label{int}
	&& \frac{1}{F} \sum_{j \neq A, B} \hspace{-0.67 cm} \int \prod \frac{d^3 p_j}{(2 \pi)^3 2 E_j} \overline{ |\hat{O} |^2} (p_j, P_H=p_A+p_B, k=0) \nonumber \\
	&& \times (2 \pi)^4 \delta^{(4)}(P_{intial}-\sum_{j \neq A, B} p_j-P_H).
\end{eqnarray}
Eq. (\ref{int}) is exactly the differential cross section of the bound state $H(A,B)$ divided by $|\Psi(0)|^2$.

There are several very basic facts supporting the extrapolation.
First of all, the amplitude and cross section are  analytical in phase space. Any practical generator should reproduce this property, and any ultraviolet divergence is not present. Secondly, the study of strong interaction is complex because of the SU(3) non-Abelian interaction, but its simulation has one simplicity: All particles taking part in the strong interactions are massive, which eliminates the infrared singularities.

Here for simplicity, we only consider the S-wave case. We take two examples: $K\bar{K}$ and $D^*N$ which are to be discussed in the next section.

\end{multicols}
	
\begin{figure}[htb]
	\centering
	\begin{tabular}{cc}
		\scalebox{0.2}{\includegraphics{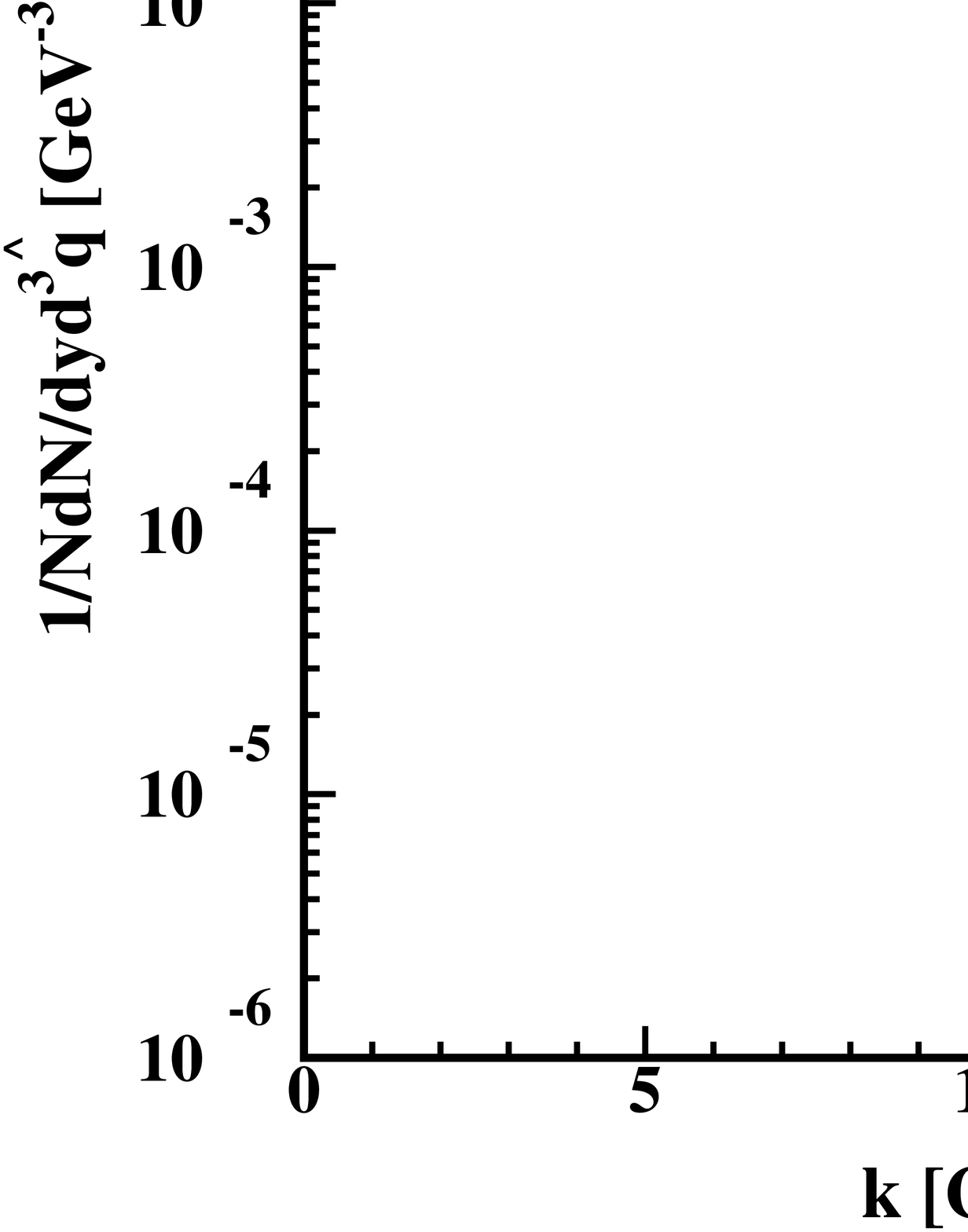}}~~&~~
		\scalebox{0.2}{\includegraphics{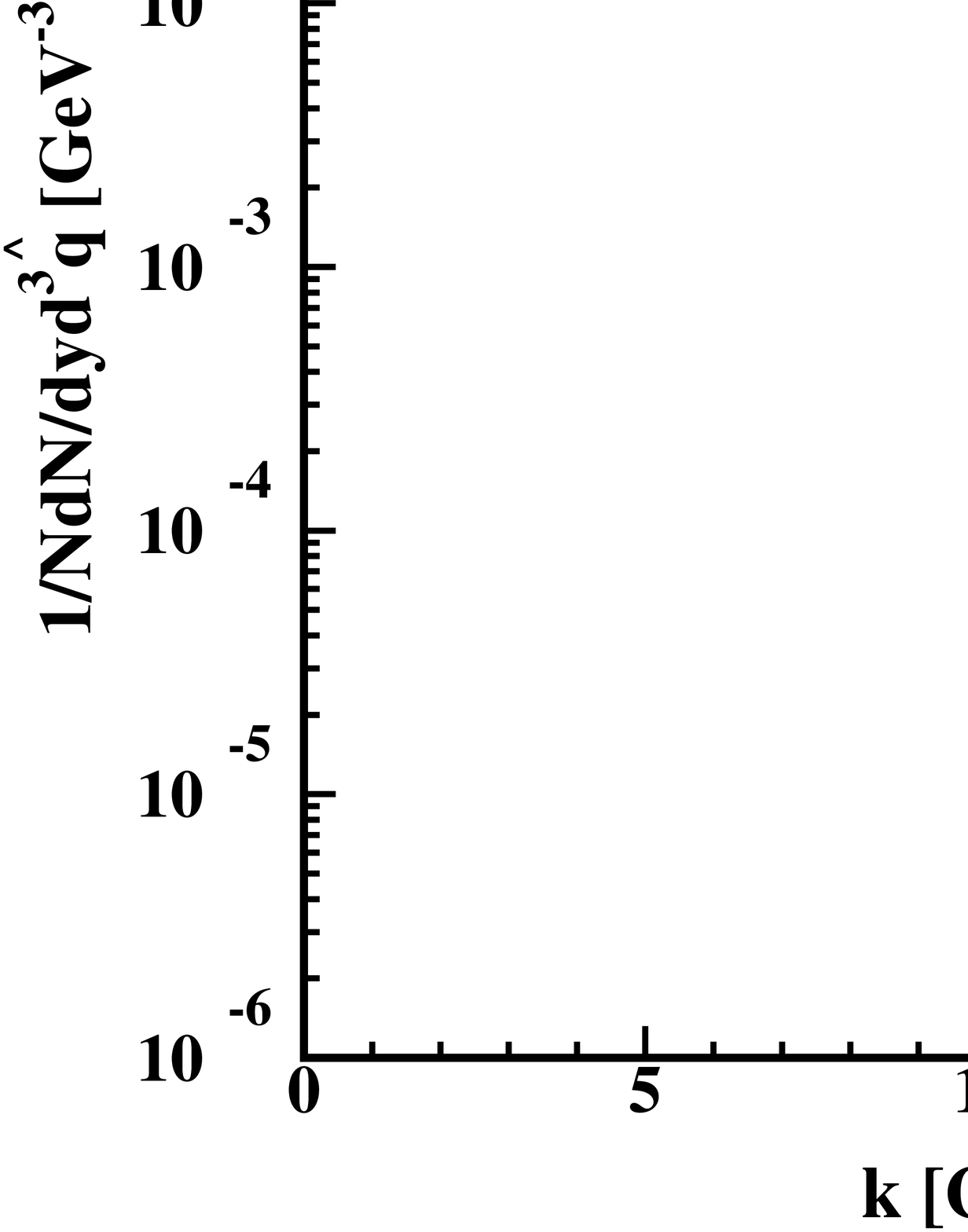}}\\
		{\scriptsize (a)}&{\scriptsize (b)}
	\end{tabular}
	\caption{
		The distribution of 3-relative momentum $k$ between the two ingredient particles: (a) $K\bar{K}$; (b) $D^*N$.
	}
	\label{qhat}
\end{figure}

\begin{multicols}{2}

Fig. \ref{qhat} shows the distribution $\frac{1}{N} \frac{d N}{d y d^3 \hat{q}}$ of these two pairs, at rapidity $y \sim 0$. These distributions are quite smooth, so one can get a reasonable extrapolating result, or equivalently using the average value around $k\sim 0$.

The smooth line also indicates that for the distribution, the derivative distribution around $k\sim 0$ is a relatively small value with respect to the distribution itself,
which then can be taken as higher order and neglected. This means that for the P-wave  production, the calculation can be much simplified from the numerical calculation of derivative.

Related with this extrapolation, we would like to address that, this formulation we employ here is  well factorized. The  NR wave function is universal. However,  this factorization is not  a priori  correct, since  we do no have an (at least effective) field theory of the hadron production. So one has no way to prove the factorization.  We can only use the generator as a numerical effective theory, itself is finite anyway.
The validity of the factorization can only be checked  for certain particle, for various  processes and/or energies (see discussion in section 4).
This depends on the concrete production mechanism and the structure of the relevant particles.
This also can be seen as a benchmark for the study, for higher order corrections.
As shown from \cite{Aaltonen:2009vj, Abulencia:2006ma, Kerzel:2006ks, Abazov:2004kp, Aaij:2011sn, Chatrchyan:2013cld, Kordas:2016ebh, Bignamini:2009sk, Artoisenet:2009wk, Bignamini:2009fn, Artoisenet:2010uu, Guo:2013ufa, Guerrieri:2014gfa}, there is an intuitive point of view, high energy collisions tend to produce pairs of hadrons with a very high relative momentum. This is indeed one of the naive points against the production of hadron molecules at
these facilities. At the same time, the detailed investigation on the dependence of the range of $q$ (or $\hat{q}$), though depending on the wave function/model of the particle such as X(3872), can give more information on the whole picture.

\begin{figure}[htb]
	\centering
	\begin{tabular}{cc}
		\scalebox{0.2}{\includegraphics{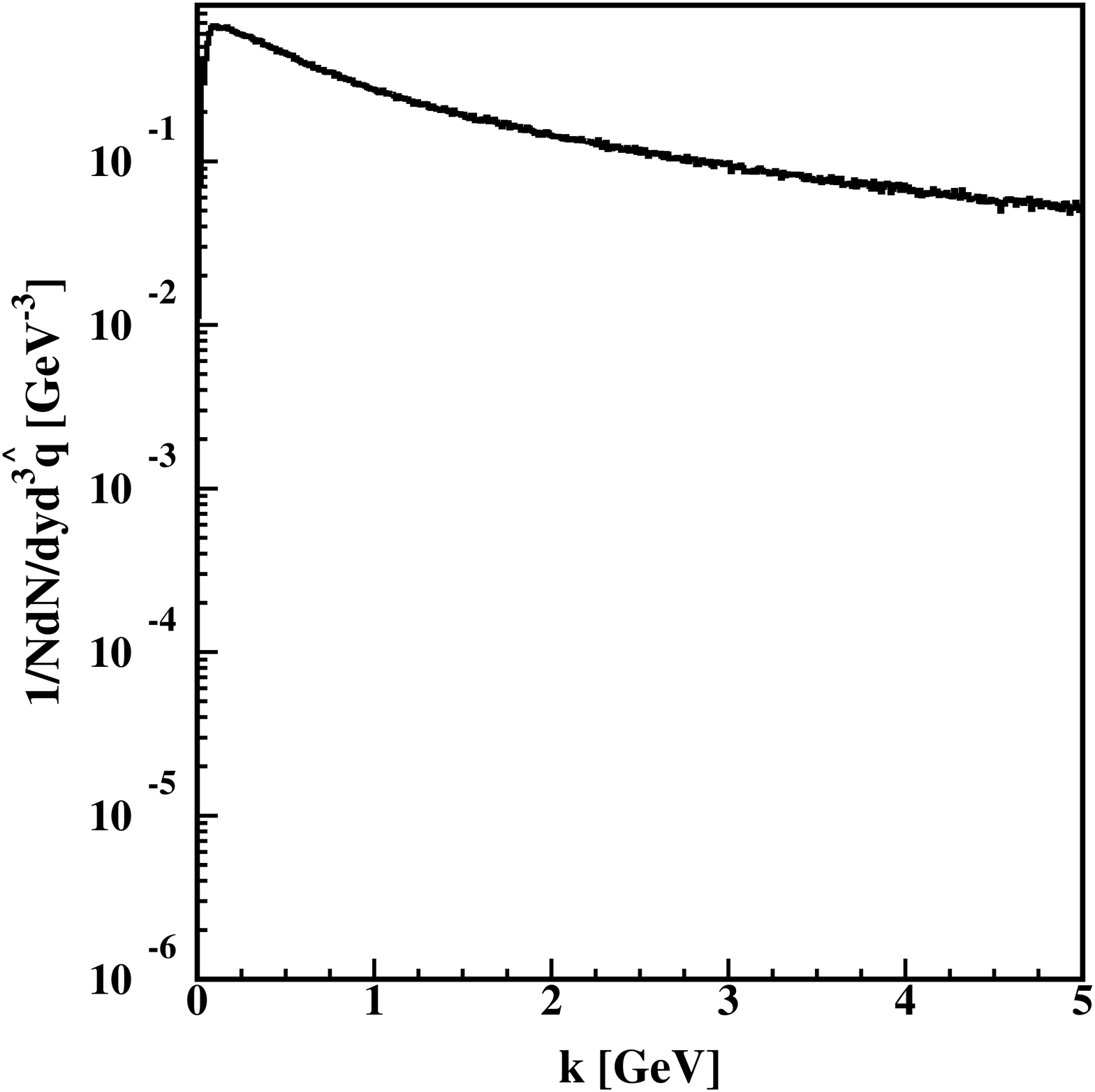}}~~&~~
		\scalebox{0.2}{\includegraphics{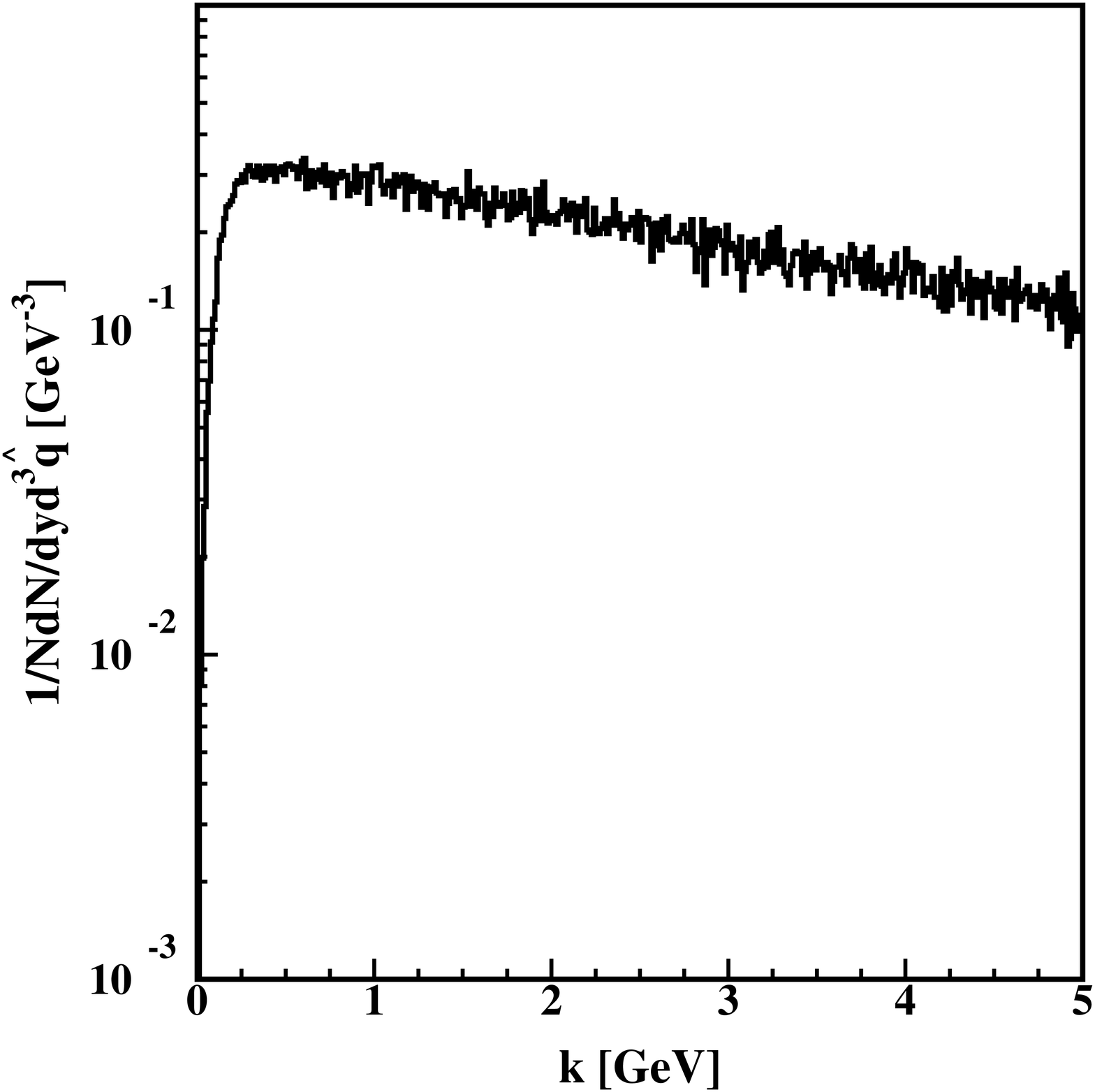}}\\
		{\scriptsize (a)}&{\scriptsize (b)}
	\end{tabular}
	\caption{
		The distribution of 3-relative momentum $k$ between the two ingredient particles at a quite small energy resolution $\delta E = 10$ MeV: (a) $K\bar{K}$; (b) $D^*N$.
	}
	\label{smare}
\end{figure}

The analysis on the factorization also leads to a more subtle consideration, {\it i.e.}, the concrete and detailed  wave function  depends on the energy and momentum resolution.  From the most practical view point, this energy and momentum resolution is determined by the experiment, especially the two hadron correlation data which is yet not available at high energy hadronic collisions. One the other hand, it is  also determined by the interplay between the high energy  hadron production mechanism and the static structures of the hadron as well as the hadron molecule.  The latter is a more deep problem and determines to what scale and energy resolution
our factorized formulation get the best validity. Here we only give another extrapolation at a quite small energy resolution for reference (see Fig. \ref{smare}). In the following
analysis we employ the former ones with `coarse resolution', which should be more consistent with the high energy processes.

\end{multicols}
	
\begin{multicols}{2}

\section{Numerical results and discussions}\label{sec4}

\end{multicols}
	
\begin{table}[htb]
	\label{bslist}
	\caption{Hadron-hadron states and related molecules.}
	\begin{center}
		\begin{tabular}{c c c}
			\hline
			~~~Hadron-hadron state~~~&~~~Related molecules~~~&~~~References for $\Psi(0)$~~~\\
			\hline
			$K\bar{K}$          &$f_0(980)$, $a_0(980)$  &  \cite{Zhang:2001sb} \\
			%            $N\bar{N}$          &$X(1835)$, $X(1860)$   & ???\\
			%            $\Delta\Delta$      &$d^*(2380)$ &   ??? \\
			$\Lambda\bar{\Lambda}$&$Y(2175)$, $\eta(2225)$&\cite{Zhao:2013ffn}\\			
			$D\bar{D}^*$        &$X(3872)$  &  \cite{Liu:2008fh,Li:2012ss,Zhao:2014gqa} \\
			&$Z_c(3900)$&  \cite{Zhao:2014gqa}         \\
			$D_s^*\bar{D}_s^*$  &$Y(4140)$  &  \cite{Liu:2009ei}\\
			$B\bar{B}^*$        &$Z_b(10610)$& \cite{Zhao:2014gqa}         \\
			$D^*\bar{D}^*$      &$Z_c(4020)$, $Z_c(4025)$  &  \cite{Zhao:2015mga} \\
			$B^*\bar{B}^*$      &$Z_b(10650)$  &  \cite{Zhao:2015mga} \\
			$D^{*0}p$           &$\Lambda_c(2940)$    &      \cite{He:2010zq} \\
			$\Lambda_c\bar{\Lambda}_c$  &$Y(4260)$, $Y(4360)$&\cite{Chen:2011cta}\\
			$\Lambda_b\bar{\Lambda}_b$  &$Y(10890)$   &    \cite{Chen:2011cta}\\
			$\Sigma_c\bar{D}^*$ &$P_c(4380)$       & \cite{Chen:2015loa} \\
			$\Sigma_c^*\bar{D}^*$ &$P_c(4450)$    & \cite{Chen:2015loa} \\
			$BK$   &X(5568)  &  \cite{Jin:2016cpv} \\
			$DK$   &$X_c$  &  \cite{Jin:2016cpv} \\
			\hline
		\end{tabular}
	\end{center}
\end{table}

\begin{multicols}{2}
%%%%%%%%%%%%%%%%%%%%%%%%%%%%%%%%%%%%%%%%%%%%%%%%%%%%%%%%%

\begin{figure}[htb]
	\centering
	\begin{tabular}{cc}
		\includegraphics[scale=0.2]{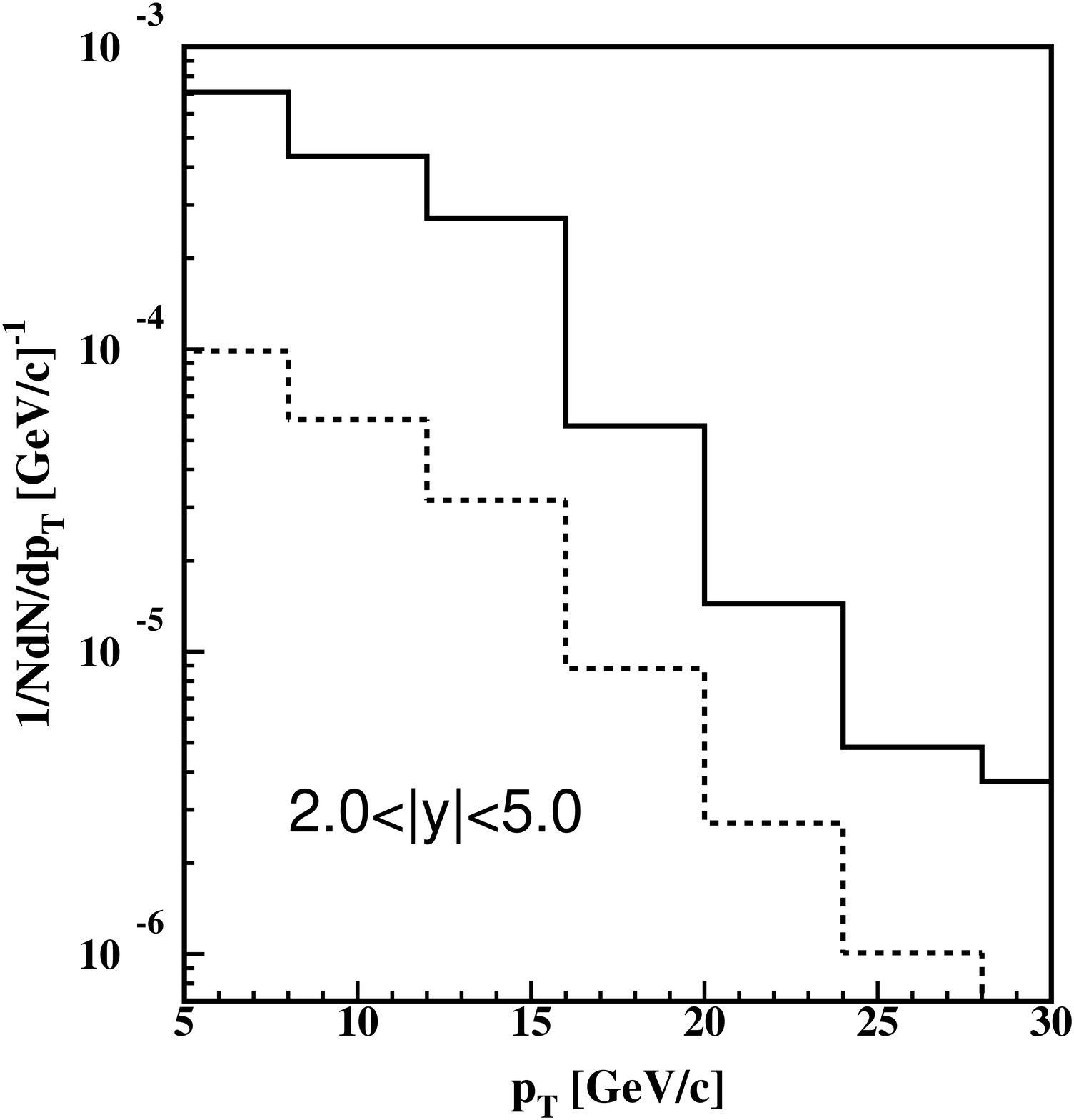}&
		\includegraphics[scale=0.2]{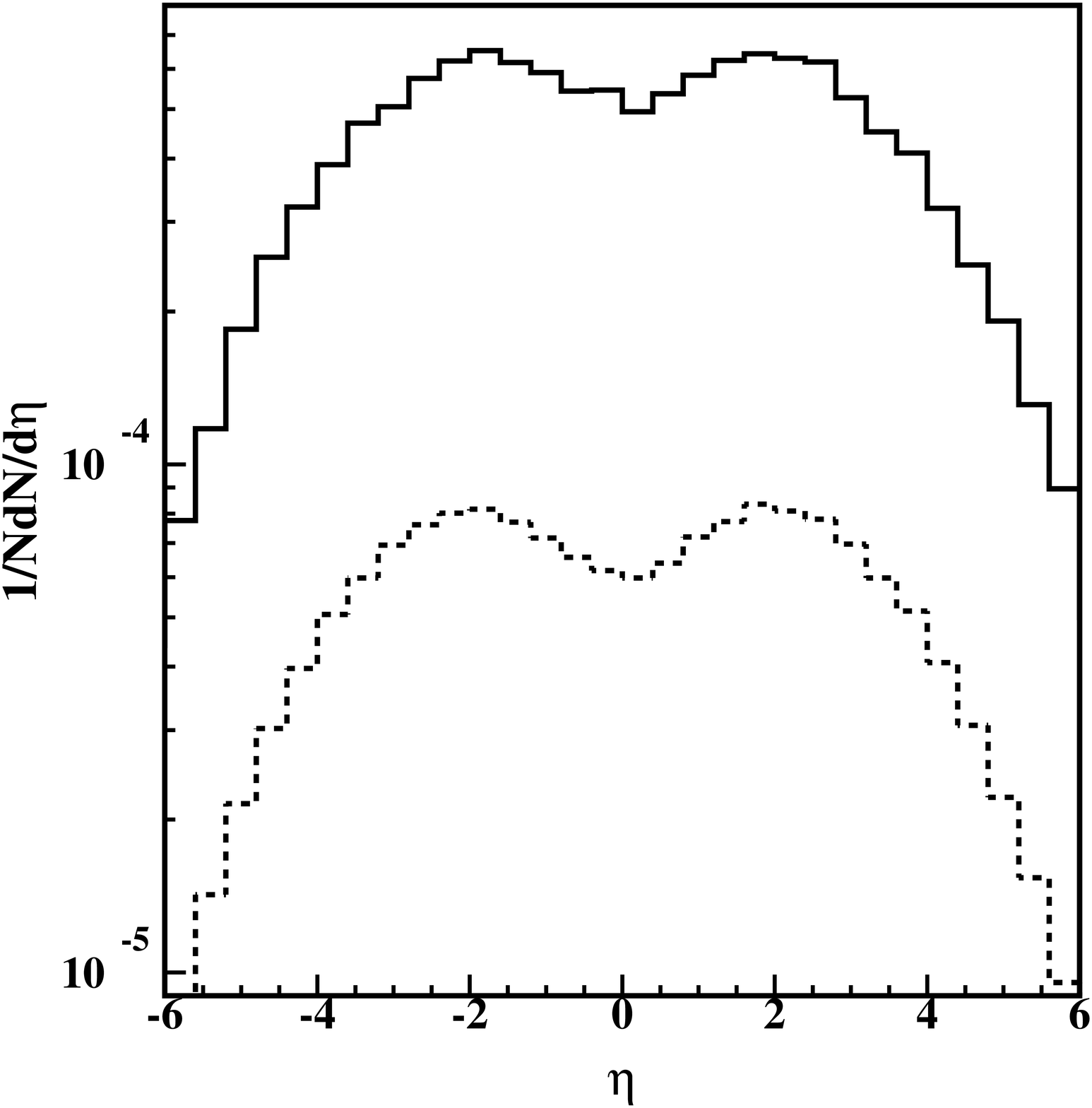}\\
		{\scriptsize (a)}&{\scriptsize (b)}
	\end{tabular}
	\caption{
		(a) Transverse momentum distributions,
		(b) Pseudo-rapidity distributions ($p_T>$ 5 GeV/c), for proton-proton collisions at $\sqrt{s}=8$ TeV. The dashed line is for X(5568), and the solid is for $B_s$.}
	\label{LHCb5568}
\end{figure}

In the above section, we have studied the production of the ingredient particle pair of $A$ and $B$. In this section, without the explicit calculation of $|\Psi(0)|^2$, we  investigate the rapidity and transverse momentum distributions (see Fig. \ref{LHCb5568}, Fig. \ref{rapidity} and Fig. \ref{pt} respectively) of some possible exotic hadron states at $\sqrt{s}=8 $ TeV in $pp$ collision.

Nowadays, a lot of hadron bound states have been studied theoretically and experimentally. Some of the  bound states and the literatures relevant to their corresponding NR wave functions are listed in Table I. In general, the wave function can be obtained by solving the relevant potential model. Employing a recent example \cite{Jin:2016cpv},
we show the wave function at the origin can also be obtained by fitting the available data, and then is used to predict the production in other phase space regions, other collision energies and processes.

Recently, the D0 collaboration announced that they found the evidence of a new state  X(5568) \cite{D0:2016mwd}, with four flavors in this hadron. If it is really a particle, this  will  be  quite remarkable, as the first solid evidence of multi-quark
state directly  produced in the multi-production process of high energy collision, rather than  from hadron decays. The bottom flavor here is very decisive. Because of the bottom flavor and the mass, it is hardly possible produced from decay of a heavier hadron.
Another important fact is that the  measurement \cite{D0:2016mwd} has given the production ratio $\rho$ of the yield of X(5568) to
the yield of the  $B_s^0$ meson in two kinematic ranges, 10 $< p_T (B_s^0) <$ 15 GeV/$c$ and 15 $< p_T (B_s^0) <$ 30 GeV/$c$.
The results for $\rho$ are (9.1 $\pm$ 2.6 $\pm$ 1.6)$\%$ and (8.2 $\pm$ 2.7 $\pm$ 1.6)$\%$, respectively, with an average of (8.6 $\pm$ 1.9 $\pm$ 1.4)$\%$.
%%%%%%%%%%%%%%%%%%%%%%%%% modification %%%%%%%%%%%%%%%%%%%%%%%%%%%%%%%%
If we assume that $B^0_s \pi^{\pm}$ is the dominant decay mode of this new state, this large production rate itself first of all excludes the possibility of decay from heavier particles like $B_c$, and is difficult to be understood by various general hadronization models~\cite{Jin:2016cpv}, such as  String fragmentation model \cite{Andersson:1998tv}, cluster model \cite{Webber-cluster} and Combination model \cite{Jin:2010wg}.
By employing these general hadronization models, one can hardly raise to a larger production ratio compared with the D0 experiment, since there are no other plausible parameters to tune. So it seems that the X(5568) has a curious structure and unique production mechanism different from the other particles produced. This needs a special theoretical framework to deal with.  Furthermore, the theoretical analysis, like the kinematic distributions of the signal particles from the
decay of X(5568), are useful for various detectors. Though recently the searches of X(5568) by LHCb and CMS give negative results \cite{Aaij:2016iev,CMS:2016fvl}, one must also consider the impacts of different cuts and trigger conditions from the D0 %\cite{Bignamini:2009sk,Artoisenet:2009wk}
adopted by those detectors.

Exploring the inclusive production formulations proposed in Sections 2 and 3, we can calculate the cross section of the new $B^0_s \pi^{\pm}$ state for various collision processes as well as energies, taking it as the bound state of two hadrons (see the 14th line of the middle column of Table I). By fitting the D0 data at the collision energy 1.96 TeV within the phase space region 10 $<p_T<$ 30 GeV/$c$ and   $|\eta|<$ 3, we get the $|\Psi(0)|^2$, which is then used to predict the production of X(5568) at LHC for the whole rapidity region, since $|\Psi(0)|^2$ is independent of collision energy and phase space region.
As an example, Fig.~\ref{LHCb5568} shows the transverse momentum distributions  and
pseudo-rapidity $\eta$ ($p_T>$ 5 GeV/c) distributions for proton-proton collisions at $\sqrt{s}=8$ TeV.
All these calculations are useful to study this unconfirmed problem since yet no other collaborations report positive results.

\end{multicols}
	
\begin{figure}
	\centering
	\scalebox{0.20}{\includegraphics{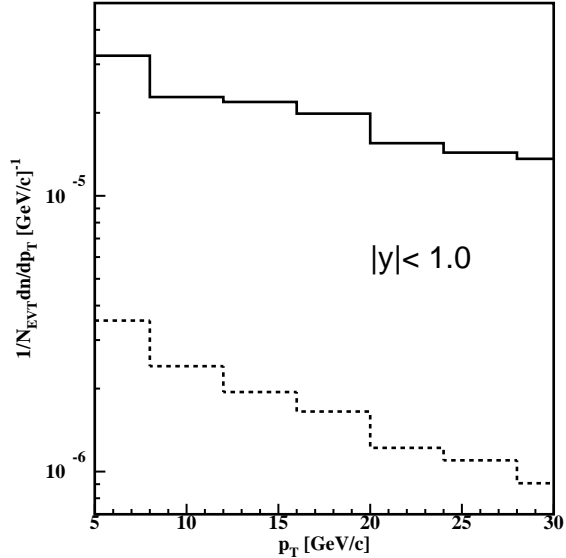}}
	\caption{Transverse momentum distributions at Tevatron. The dashed line is that for X(5568), with the best fitting of the wave function to get the
		correct $\rho$ measured by D0 collaboration. The solid line is for $B_s$ as reference.}
	\label{ptbs}
\end{figure}

%%%%%%%%%%%%%%%%%%%%%%%%%%% modification %%%%%%%%%%%%%%%%%%%%%%%%%
\begin{multicols}{2}
	
The X(5568) transverse spectrum of $p_T$ is softer than that of $B_s$, as demonstrated in Fig.~\ref{ptbs} and indicated from the experiment \cite{D0:2016mwd}.
This is from the fact that the two clusters are required to be close to each other in the phase space for combination.
Realized in the above formulations in Section 2, is the relative momentum vanishing.
So that this framework generally predicts a softer $p_T$ spectrum: The behavior looks like a single particle in small $p_T$.
However, the production rate is suppressed for large $p_T$. The reason is very clear. If we assume randomly correlations between two ingredient particles, the larger the single transverse momentum of them, the smaller the probability of the two particles with almost vanishing relative momentum, and hence much more suppressed. This is a very typical property of the NR formulations, in contrary to the fragmentation spectrum, the more massive, the harder.
For high energy and then high $p_T$ processes, the question whether the NR formulations are still valid or not is not settled yet. The quarkonium production had given some implications. A possible way is to employ the Bethe-Salpeter wave function \cite{Bhatnagar:2005vw,Bhatnagar:2009jg}. In this formulation, all possible relative momenta of the ingredients are taken into consideration.
Once the wave function shows that the probability of large relative momentum between the ingredients is vanishing, the Bethe-Salpeter description might go to the NR formulations. However, the rich Dirac structures in the Bethe-Salpeter wave functions can also introduce sounding informations.
In Refs. \cite{Bignamini:2009sk, Artoisenet:2009wk, Bignamini:2009fn, Artoisenet:2010uu, Guerrieri:2014gfa}, the phase space wave function is directly used, with a certain significant momentum region. In these kind of models, the bound states can have a harder spectra.

Furthermore, both charm and bottom quarks are heavy. Relevant processes can be calculated with perturbative QCD in the exactly same way once the different values of the masses are taken into account. So, when one replaces the bottom quark by the charm quark in the X(5568) and assumes that the structure and the production mechanism do not change, the production of the `new $D^{\pm}_s \pi^{\pm}$ state' ($X_c$) can be also studied. Since the reduced mass is mainly determined by the mass of the light ingredient hadron, say, Kaon here, the wave function at the origin can be taken as the same. Therefore, the cross section of $X_c$ for both Tevatron and LHC should be completely determined. We illustrate the rapidity and transverse momentum distributions of $X_c$ in Fig. \ref{rapidity} (a) and Fig. \ref{pt} (a). If $X_c$ exists, it is not difficult to be detected: the $D^{\pm}_s$ can be determined from the $D^{\pm}_s \to \phi \pi$ channel, by proper 3-charged particle tracks from the vertex displaced from the primary one; then this reconstructed $D^{\pm}_s$ can be combined with a proper charged particle track considered as $\pi$ from the primary vertex to give the invariant mass distribution to look for the resonance. If the $K^0_s$ is well measured, the $D^{\pm}_s$ can also be reconstructed from the 2K channel and then combined with the $\pi$ from the primary vertex. This kind of pions can eliminate the possibility that the $X_c$ is produced from the decay of bottom. Of course just by keeping or not this restriction, one can preliminarily investigate $X_c$ from multi-production or from weak decay.

All the above discussions are based on the factorization formulas, {\it i.e.}, the mechanism of the multiproduction in hard process and the wave function of a certain bound state are universal. 
The only difference between the $pp$ and $\ppbar$ machines lies in the center of mass energy and the parton distribution functions. Comparing Fig. 4 and 5, the transverse momentum distributions can be different because of the difference on center of mass energy, parton distribution functions and rapidity region. But the behavior of X(5568) are quite similar as that of $B_s$. Fatal violation of the factorization will lead to more difference of the X(5568) production between these two kinds of collisions.

To further demonstrate this framework, we make a global analysis of the prompt X(3872) data. 
Contrary to the case of X(5568), it is a well-established particle \cite{Olsen:2004fp, Braun:2014rsa, Abulencia:2006ma, Chatrchyan:2013cld}, but without definite evidence for its structure \cite{Bignamini:2009sk, Artoisenet:2009wk, Bignamini:2009fn, Artoisenet:2010uu, Guerrieri:2014gfa}. However, as an exotic hadron state, its high production rate is also remarkable.
Here, taking it as hadron bound state as shown in Table I, employ the available data
from CDF and CMS \cite{Abulencia:2006ma, Chatrchyan:2013cld}, and the event generator, we calculate the following value:
\begin{equation} \label{ext1}
	R=\int_{-y_0}^{y_0} \int_{p_{T0}} \frac{1}{N_{<\Psi(2S)>}} \frac{d N_{X(3872)}}{dy d^2p_T d^3 \hat{q}}|_{\hat{q}=0}/\sigma_{X}^{exp}.
\end{equation}
Here $y$ and $p_T$ are the rapidity and transverse momentum of bound state respectively.   $y_0=0.6$ and $p_{T0}=5$ GeV/c for CDF; $y_0=1.2$ and $p_{T0}=10$ GeV/c for CMS.  The $N_{\Psi(2S)}$ is the number of $\Psi(2S)$ for the corresponding collision. The $\frac{d N_{X(3872)}}{dy d^2p_T d^3 \hat{q}}$ is calculated by event generator. The $\sigma_{X}^{exp}$ is the prompt production cross section of X(3872) of experimental data, and here we use $\sigma_{X}^{exp}=3.1$ nb for CDF and $\sigma_{X}^{exp}=1.06$ nb for CMS.
We find  $R_{CMS} / R_{CDF} $ is of order of 1, the universality of the wave function and  hence factorization is approved. 
  It is considered in some literatures \cite{Liu:2008fh,Li:2012ss,Zhao:2014gqa} that $Z_c$ shares some similarity with X(3872). Because $Z_c$ is apparently a tetraquark, its production rate is relatively small in multiproduction process, as has been discussed in Section 1. If it is effectively dealt with as hadron molecule, its wave function at origin should be much smaller than that of X(3872). The similar investigation can also be applied to $Z_b$.

The distributions of the rapidity and the transverse momentum are shown in Fig. \ref{rapidity} (with $p_T>$ 5 GeV/c) and \ref{pt} for the corresponding bound states listed in Table I. It is obvious that all the bound states have the similar shape and property, like $X(5568)$ and $X_c$ \cite{Jin:2016cpv}.
Fig.~\ref{rapidity} and \ref{pt} indicate separately that the rapidity distribution can extend to $|y|>5$, and the transverse momentum distribution still has some detectable value when $p_T \sim 40$ GeV.

\end{multicols}

\begin{figure}[htb]
	\centering
	\begin{tabular}{cccccc}
		\scalebox{0.20}[0.18]{\includegraphics{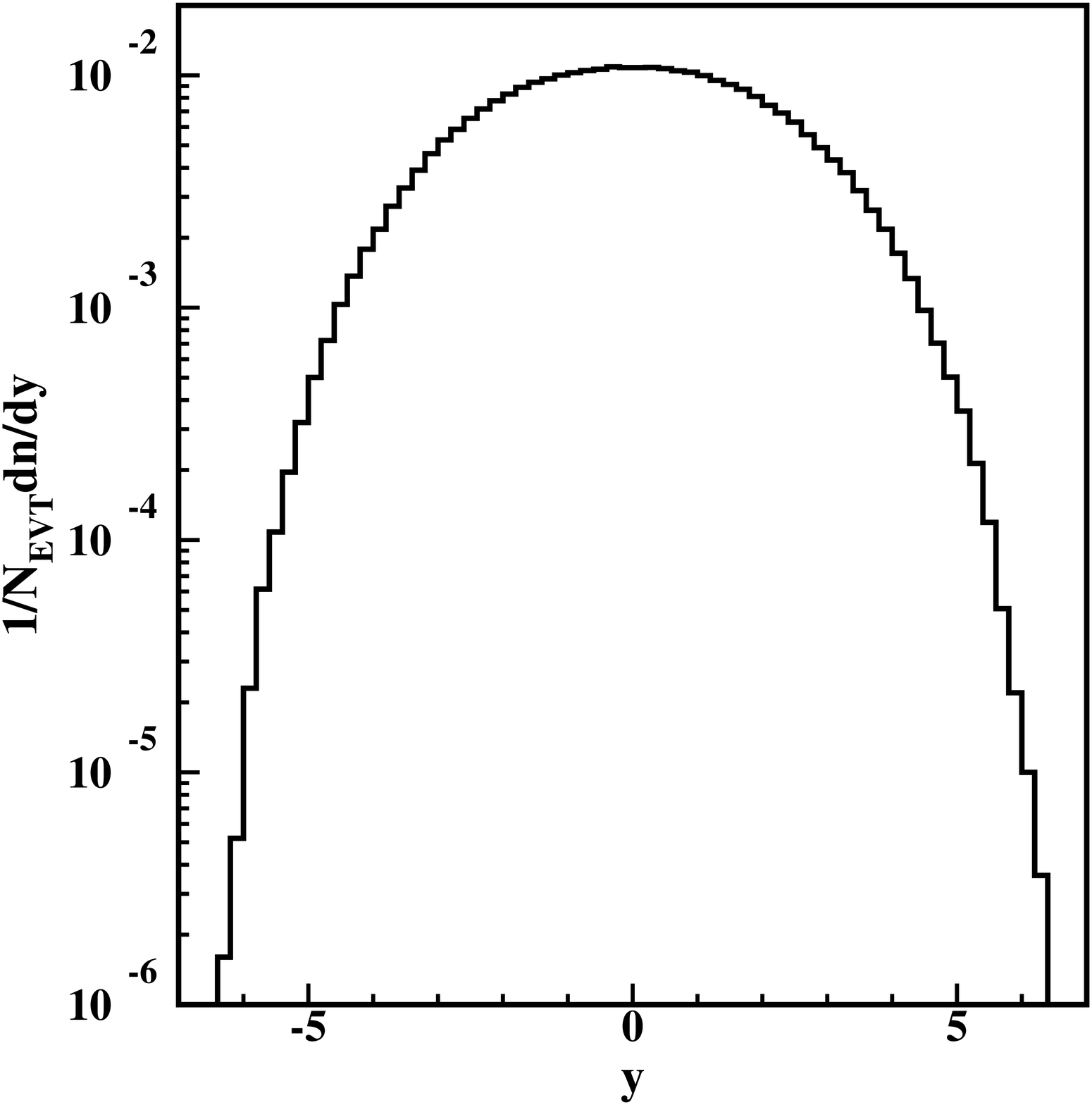}}&
		\scalebox{0.20}[0.18]{\includegraphics{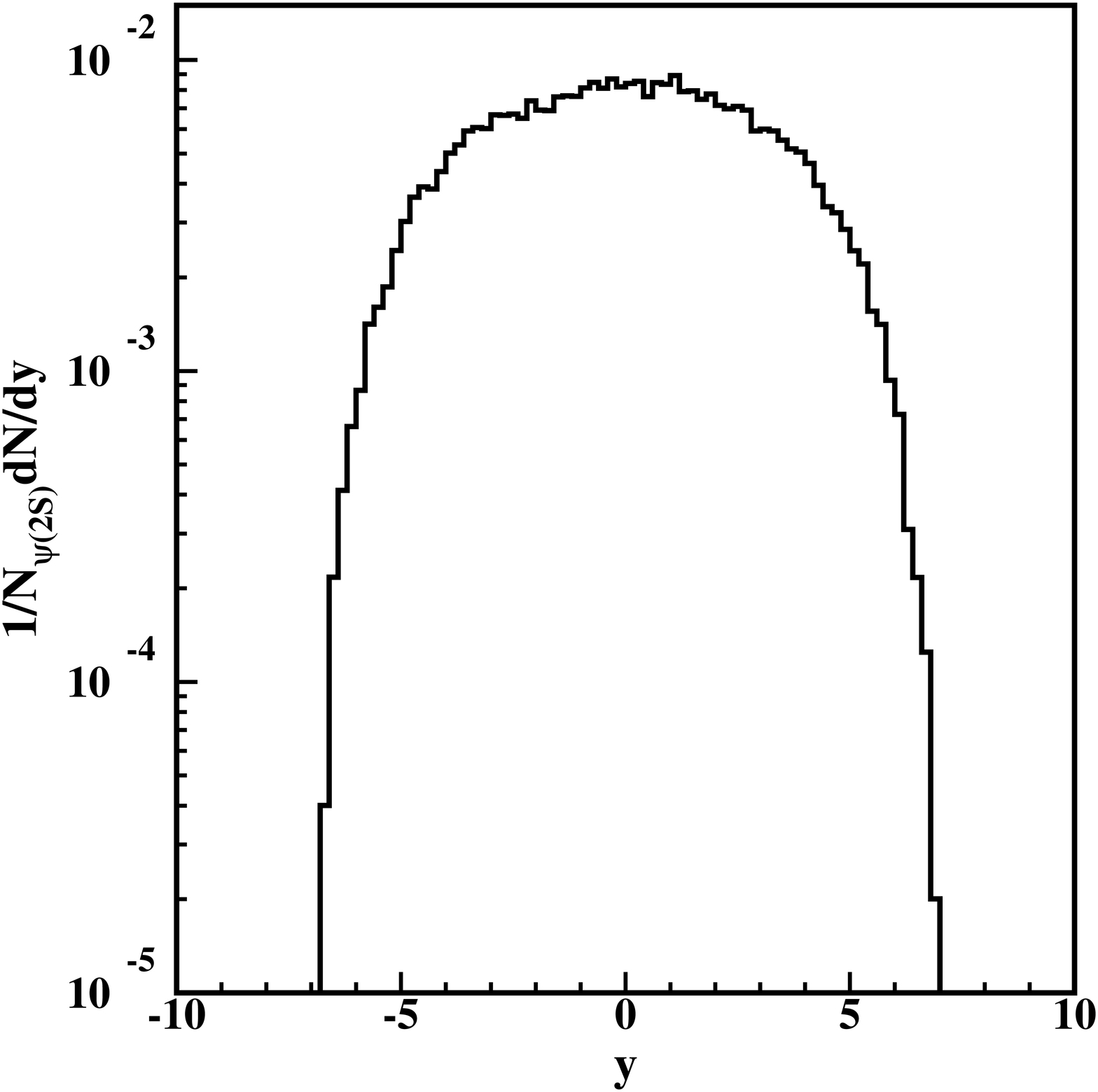}}&\\
		{\scriptsize (a)}&{\scriptsize (b)}\\
		\scalebox{0.20}[0.18]{\includegraphics{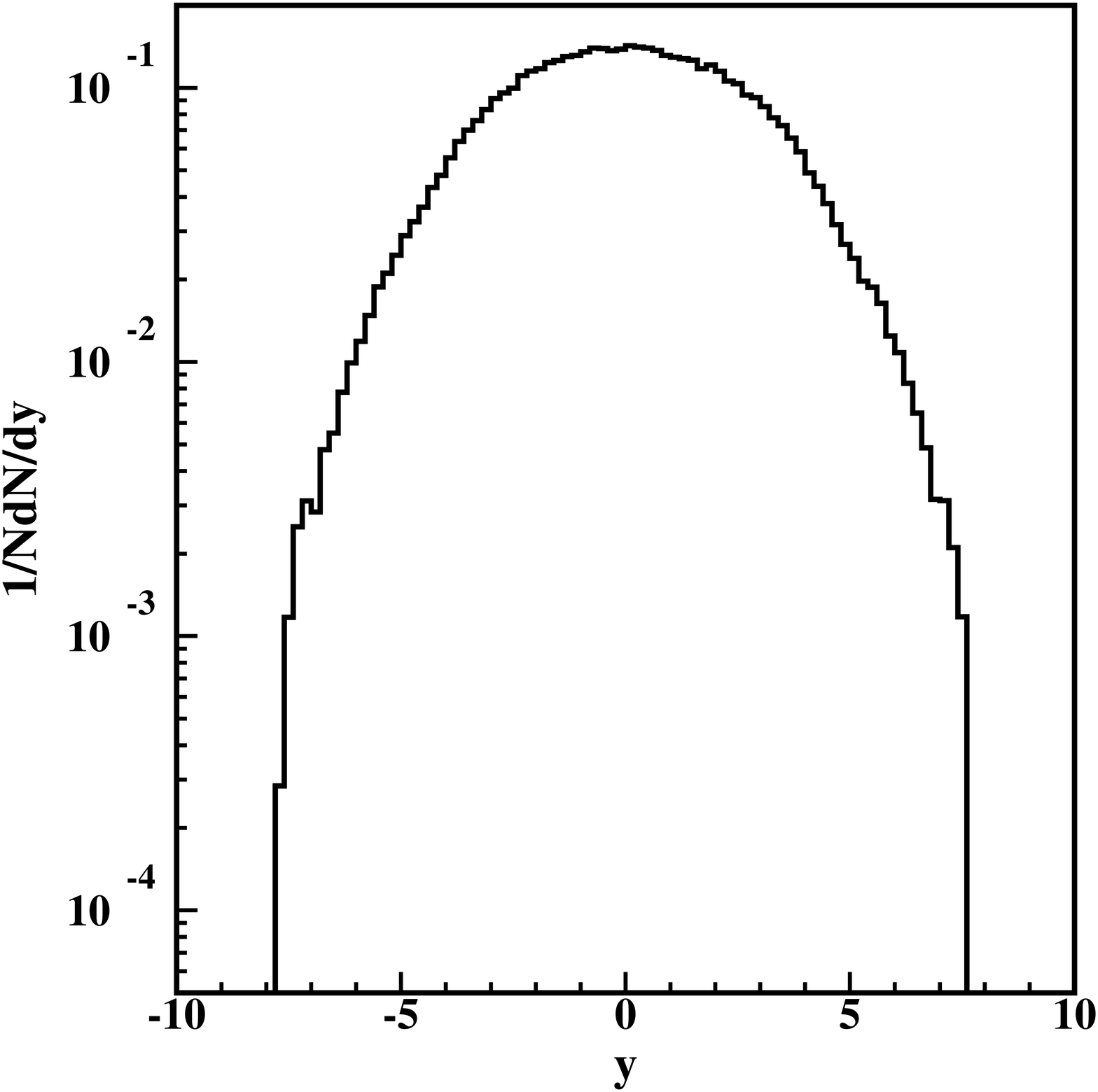}}&
		\scalebox{0.20}[0.18]{\includegraphics{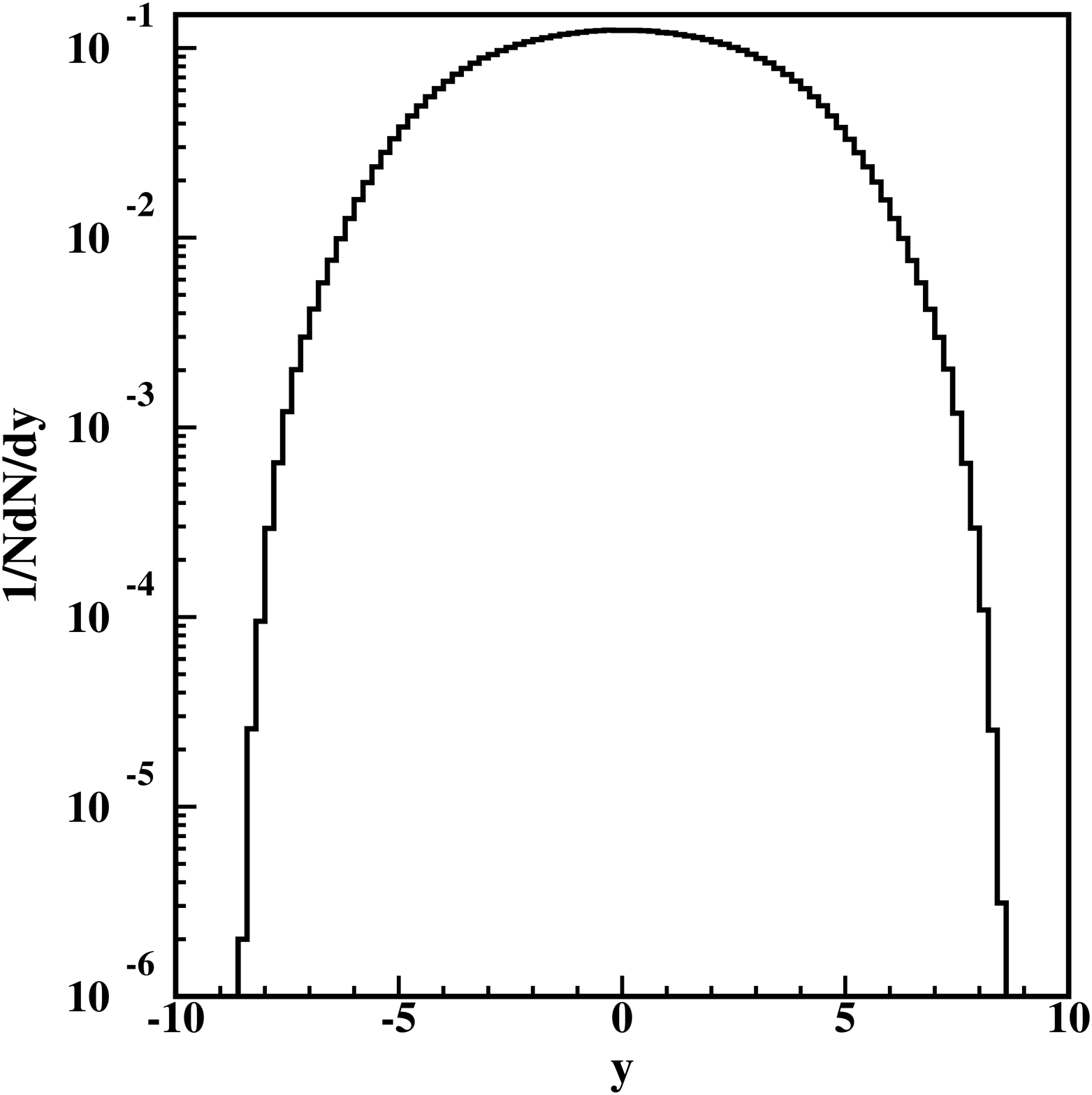}}&\\
		{\scriptsize (c)}&{\scriptsize (d)}\\
		\scalebox{0.20}[0.18]{\includegraphics{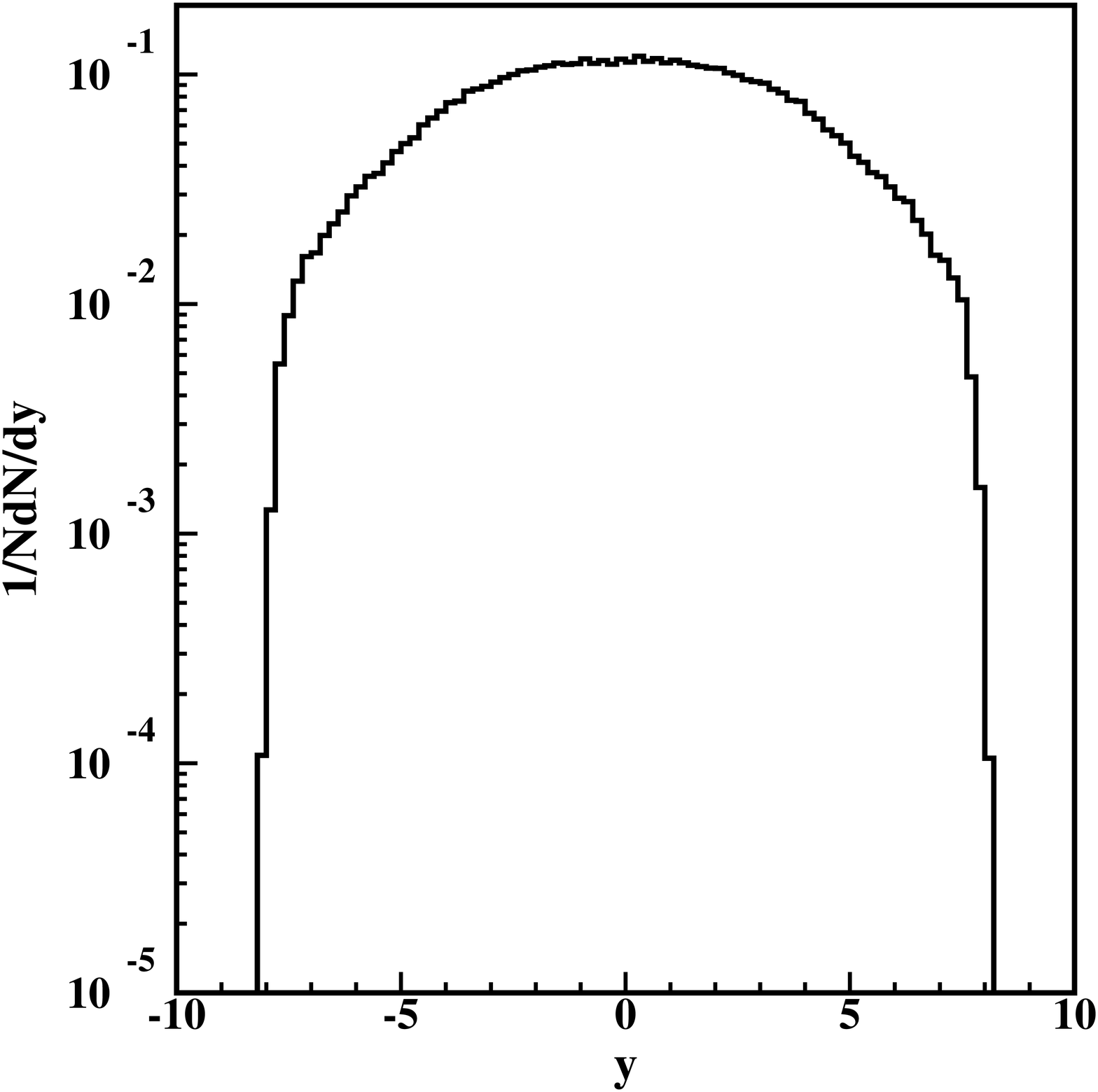}}&
		\scalebox{0.20}[0.18]{\includegraphics{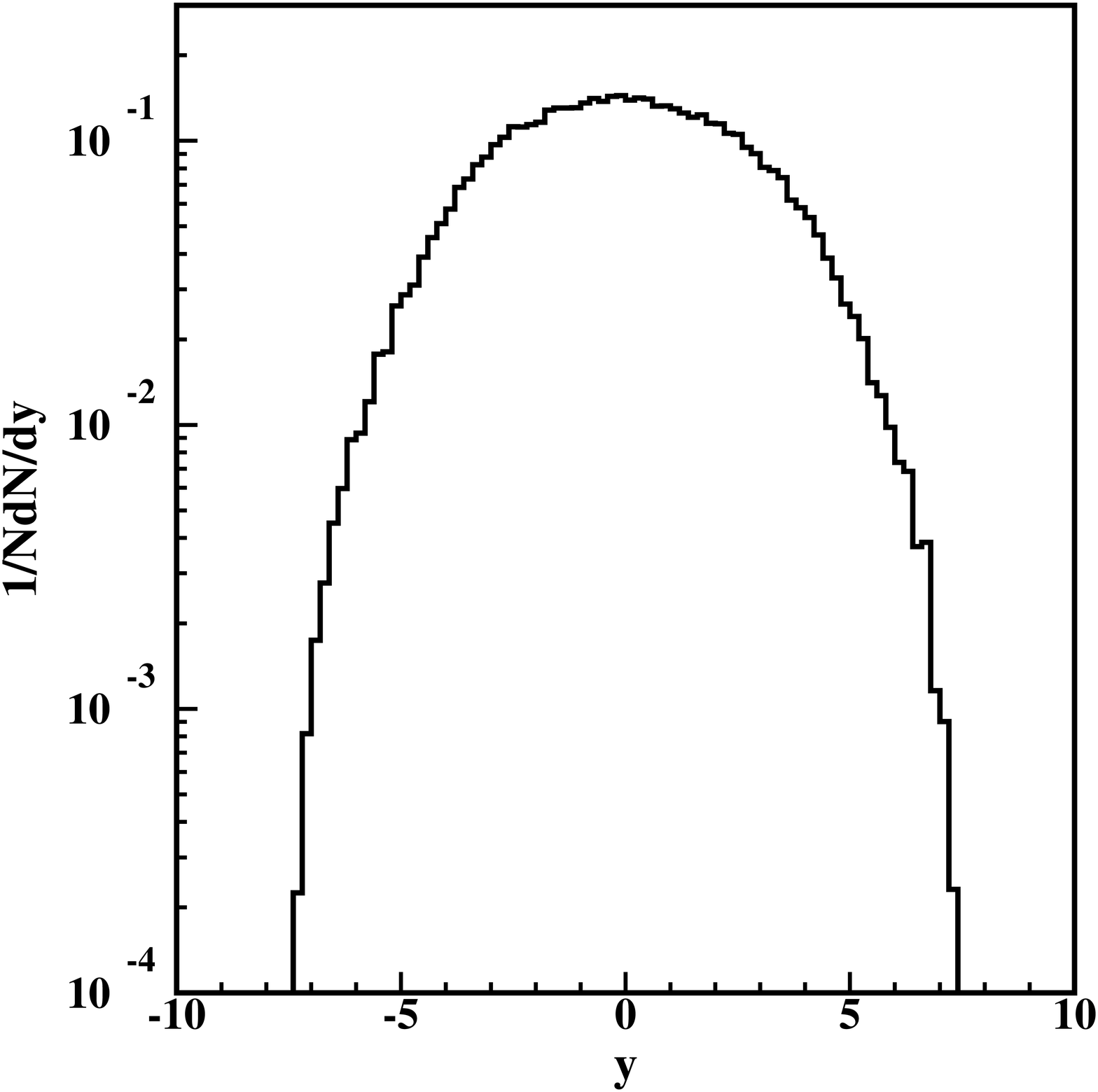}}\\
		{\scriptsize (e)}&{\scriptsize (f)}
	\end{tabular}
	\caption{
		The rapidity distributions (with $p_T>$ 5 GeV/c) for the bound states for proton-proton collisions at $\sqrt{s}=8$ TeV, (a) $X_c$: $D^{\pm}_s \pi^{\pm}$ is the production rate, (b) $X(3872)$: $D\bar{D}^*$ is the production rate with respect to $\psi(2S)$. (c) $\Lambda_c(2940)$: ($D^*N$), (d) $f_0(980)$: ($K\bar{K}$), (e) $Y(4260)$: $\Lambda_c\bar{\Lambda}_c$, (f) $P_c(4380)$: $\Sigma_c\bar{D}^*$ are normalized to one.}\label{rapidity}
\end{figure}

\begin{figure}[htb]
	\centering
	\begin{tabular}{cccccc}
		\scalebox{0.20}[0.18]{\includegraphics{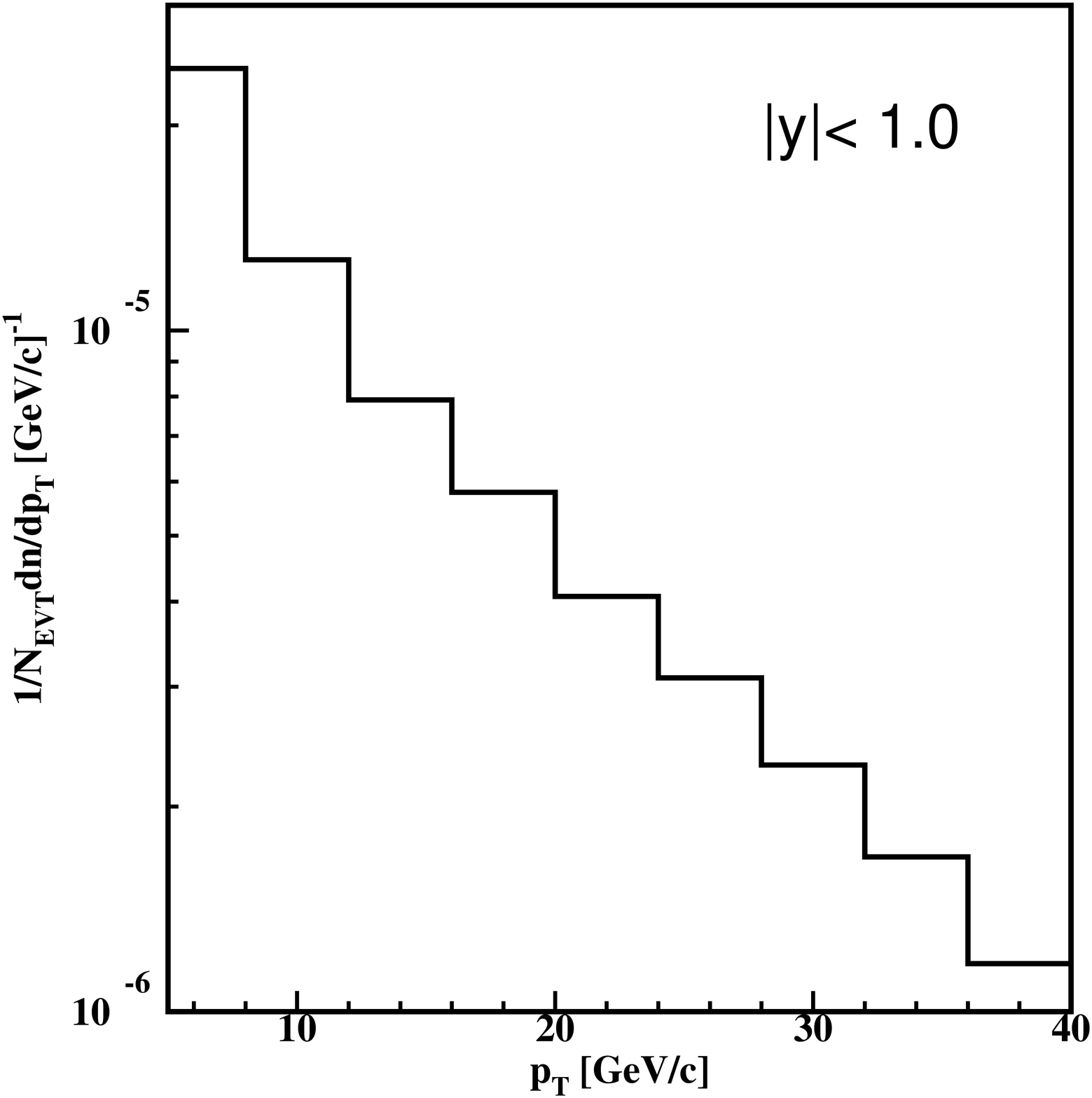}}&
		\scalebox{0.20}[0.18]{\includegraphics{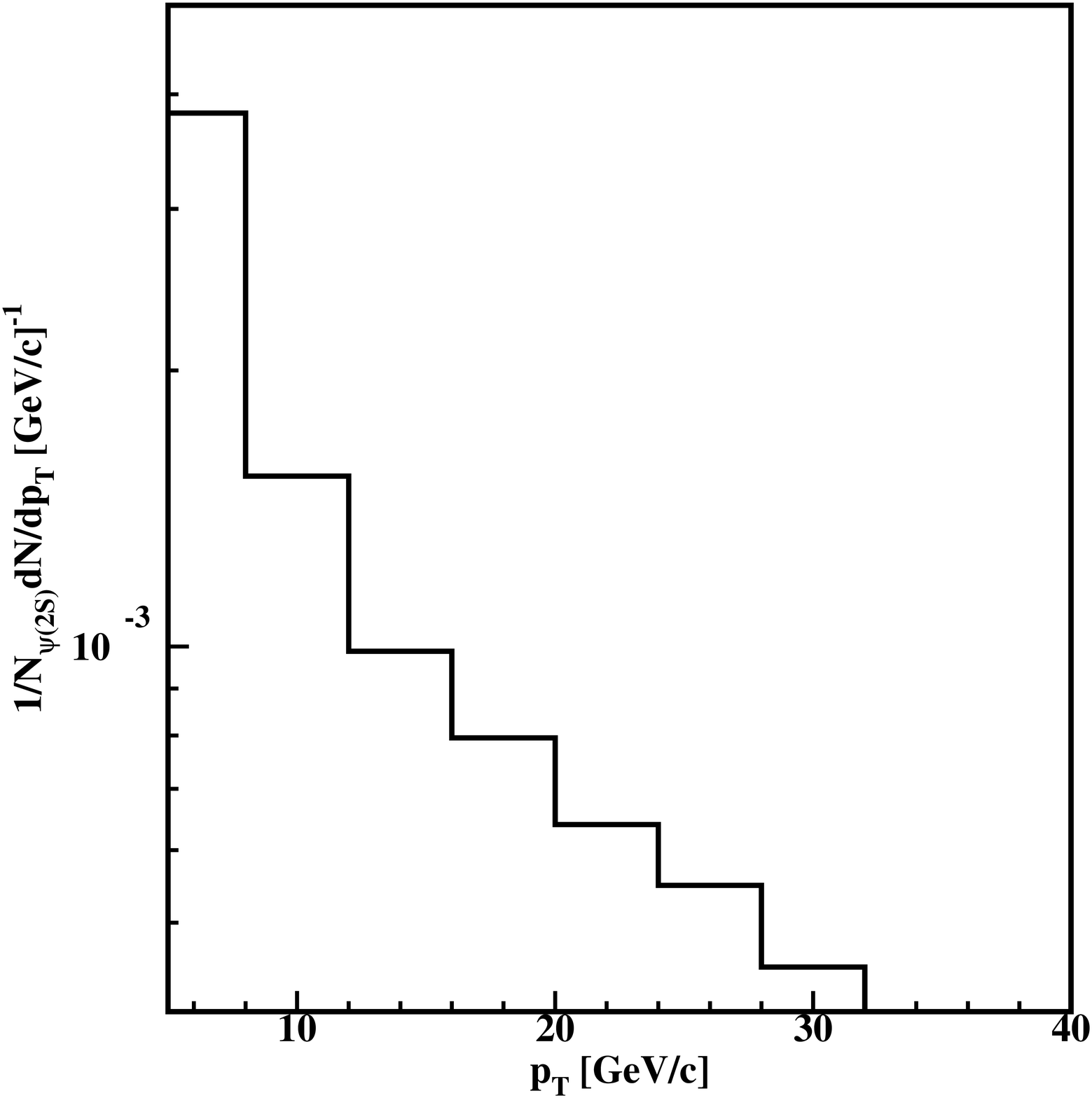}}&\\
		{\scriptsize (a)}&{\scriptsize (b)}\\
		\scalebox{0.20}[0.18]{\includegraphics{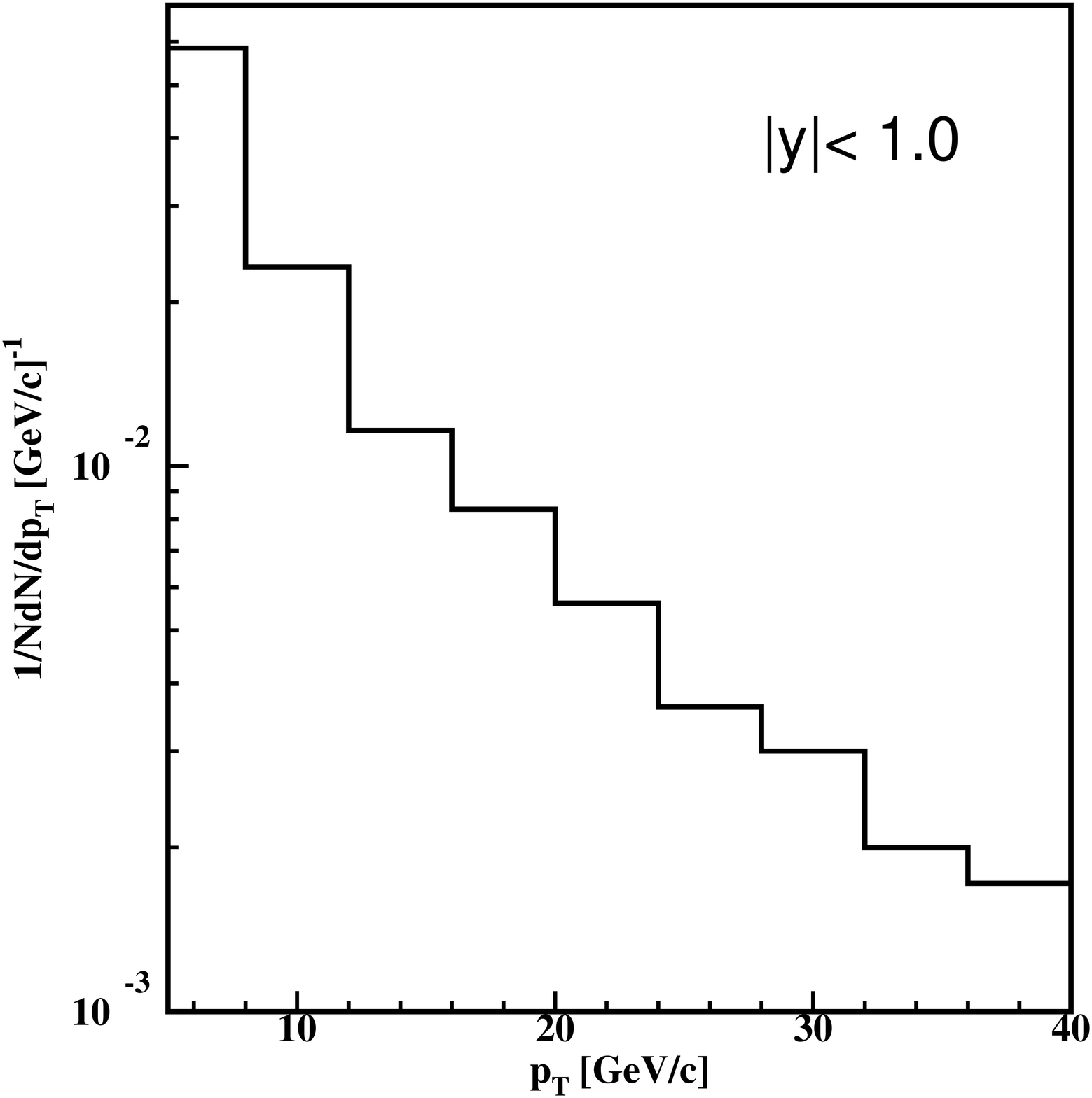}}&
		\scalebox{0.20}[0.18]{\includegraphics{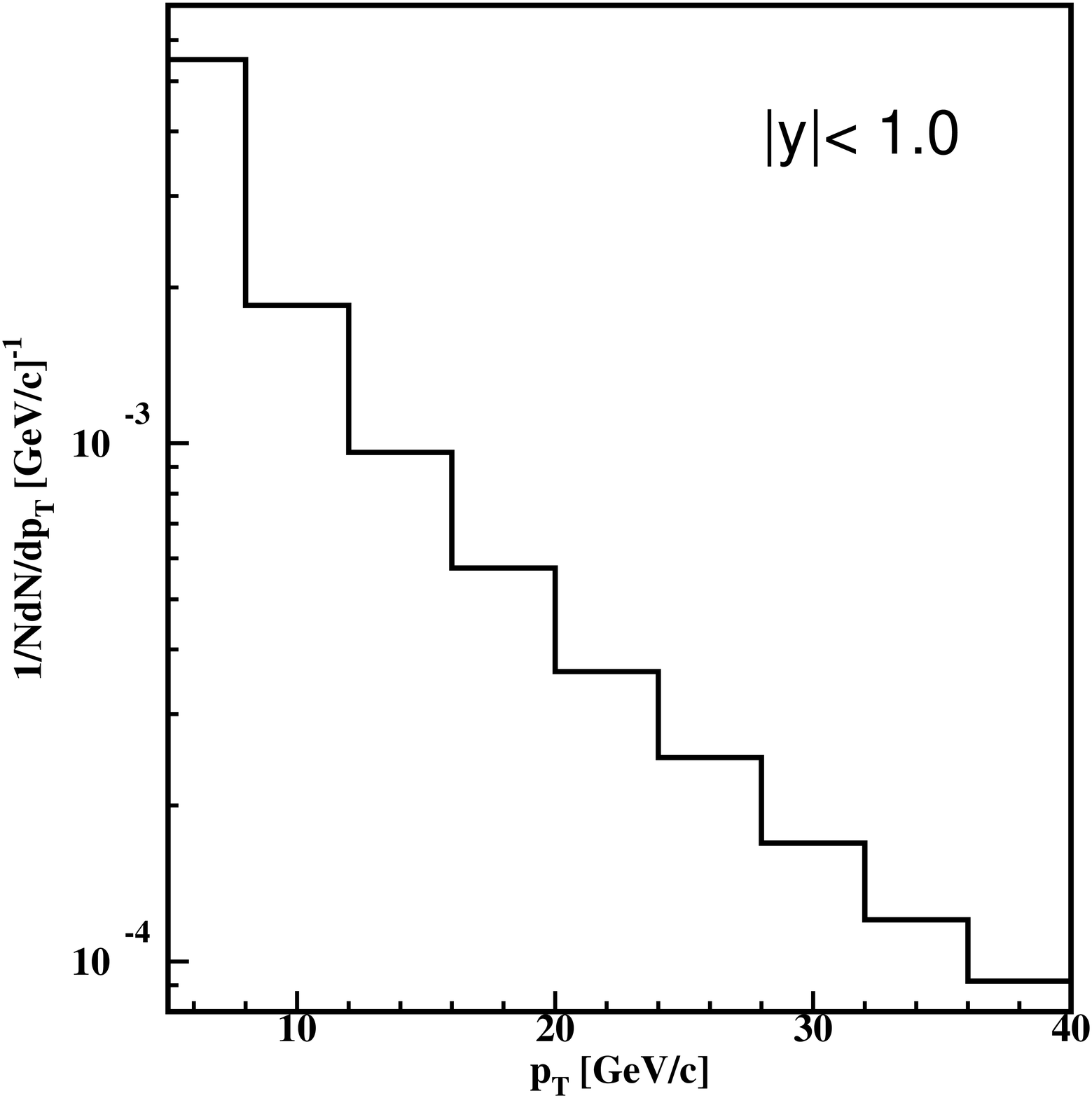}}&\\
		{\scriptsize (c)}&{\scriptsize (d)}\\
		\scalebox{0.20}[0.18]{\includegraphics{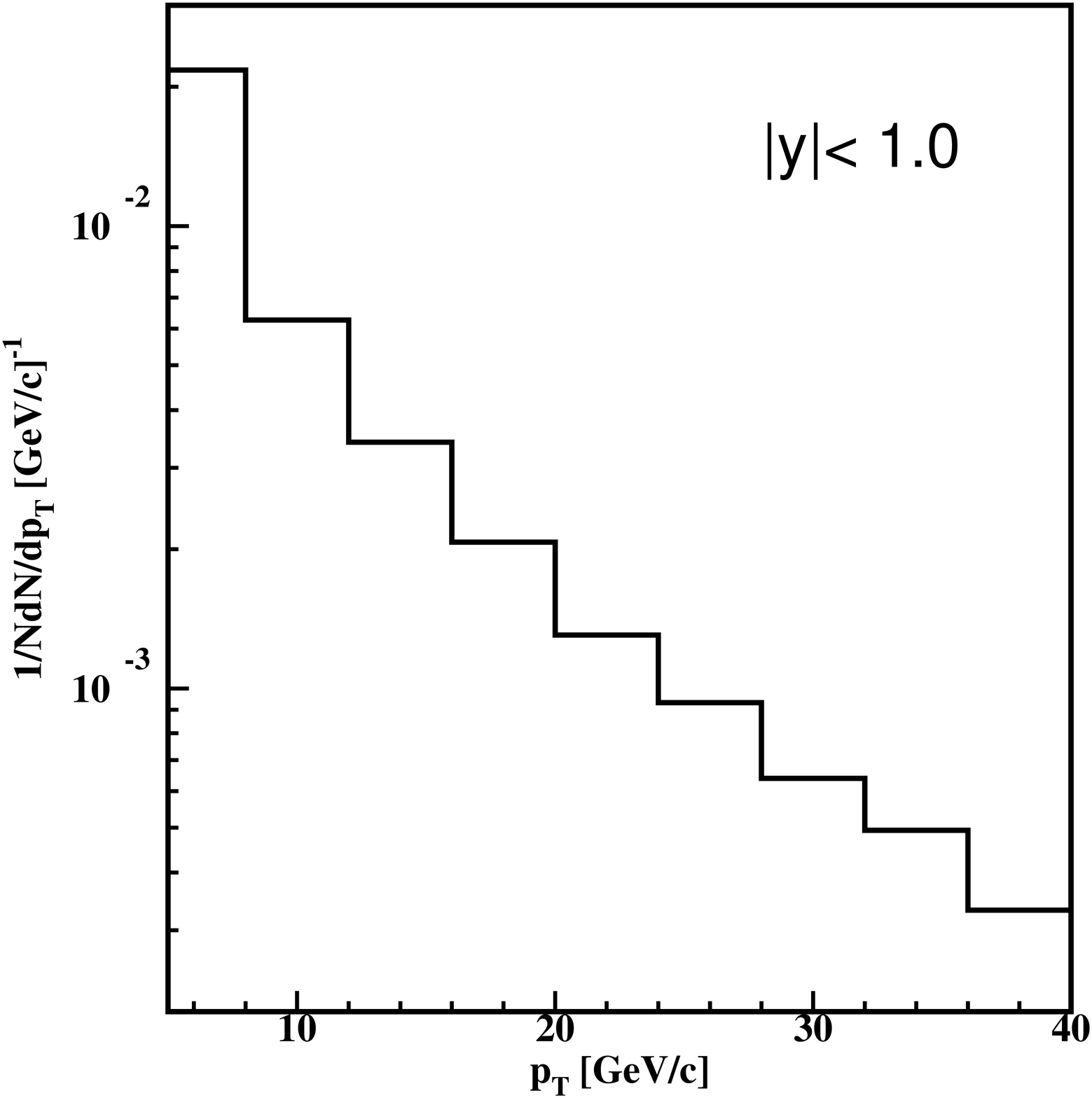}}&
		\scalebox{0.20}[0.18]{\includegraphics{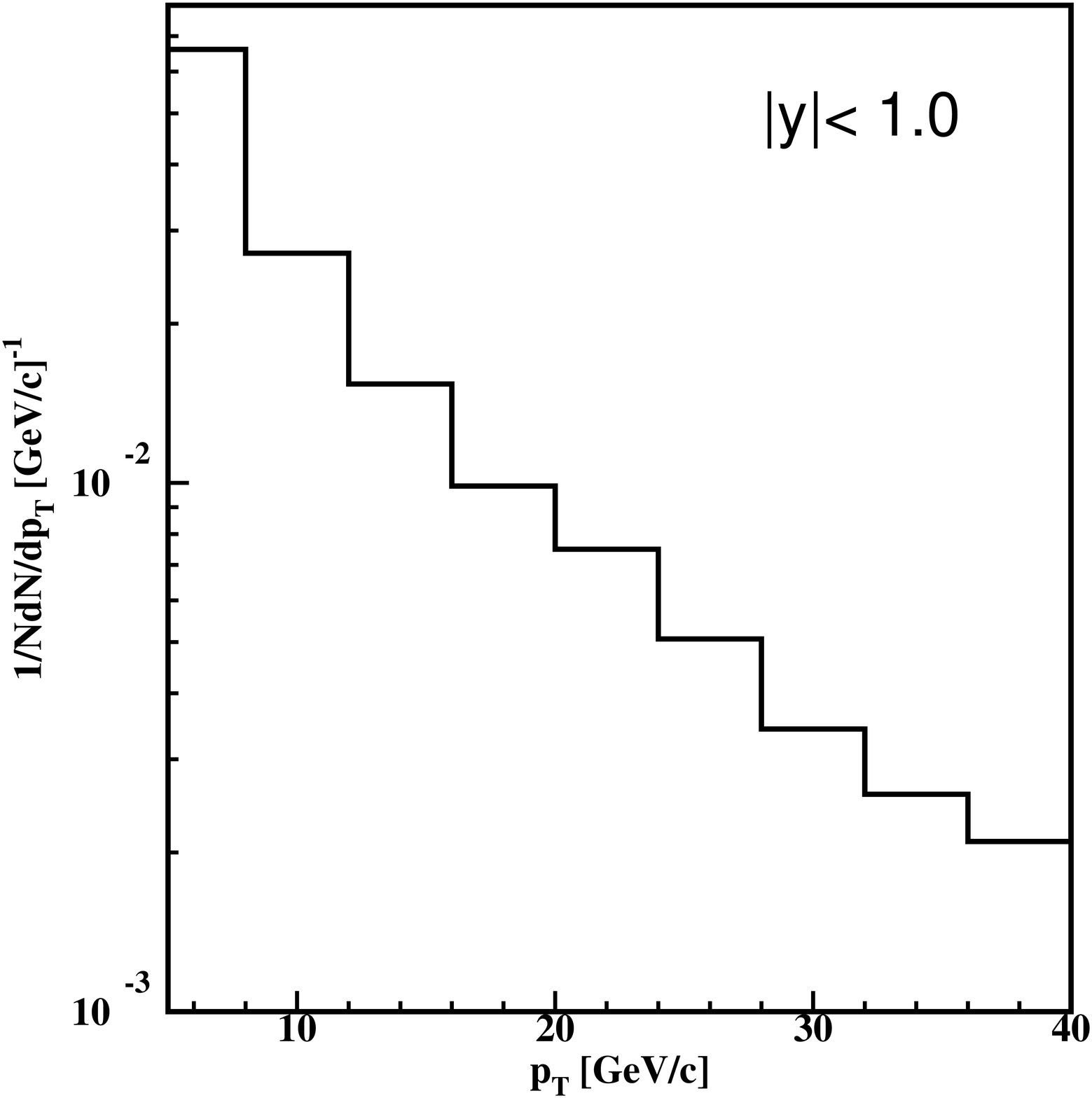}}\\
		{\scriptsize (e)}&{\scriptsize (f)}
	\end{tabular}
	\caption{
		The transverse momentum distributions for the bound states for proton-proton collisions at $\sqrt{s}=8$ TeV, (a) $X_c$: $D^{\pm}_s \pi^{\pm}$ is the production rate, (b) $X(3872)$: $D\bar{D}^*$ is the production rate with respect to $\psi(2S)$. (c) $\Lambda_c(2940)$: ($D^*N$), (d) $f_0(980)$: ($K\bar{K}$), (e) $Y(4260)$: $\Lambda_c\bar{\Lambda}_c$, (f) $P_c(4380)$: $\Sigma_c\bar{D}^*$ are normalized to one.}
	\label{pt}
\end{figure}

\begin{multicols}{2}

The above results are useful for various detectors. For example, based on our calculation, one can go further to estimate the kinematic distributions of the signal particles which are from the decay of X(5568) and can be directly detected.
In \cite{Jin:2016cpv}, we give the transverse momentum - total momentum distribution for the signal pions from the decay process $X \to B_s + \pi$ in the LHCb detector range ($2< \eta <5$). The mass difference between X(5568) and $B_s+\pi$ is  small, and the pion mass is small. These facts lead to that the produced pions are not energetic, {\it e.g.}, only around $10 \%$ of the signal pions with transverse momentum larger than 0.5 GeV/$c$
(the requirement of the relevant measurement by the LHCb Collaboration \cite{Aaij:2016iev}).
However, since the mass of the $X_c$ is around half of the $X(5568)$, it has a larger Lorentz boost factor about two times of that of X(5568) for the same momentum. This means that wherever Tevatron or LHC, in both central and large rapidity regions, the signal pions are more energetic to be detectable.

Additionally, if the present generators are well modified for the production of polarized hadrons when the polarization experimental data are sufficient, the method in this paper can also give the result of the bound state with larger spin. This is a good interplay arena between hadron polarization and exotic hadron states studies.

\acknowledgments{The authors thank all members of the Theoretical Particle Physics Group of Shandong University for their helpful discussions.}

\end{multicols}

\vspace{-1mm}
\centerline{\rule{80mm}{0.1pt}}
\vspace{2mm}

\begin{multicols}{2}

\end{multicols}

\clearpage
\end{CJK*}
\end{document}